\documentclass[twocolumn,prl,superscriptaddress]{revtex4-1}
\usepackage{amsmath,amssymb,mathrsfs}
\usepackage{natbib}
\usepackage{subfigure}
\usepackage{tabularx}
\usepackage{epsfig}
\usepackage{longtable}
\usepackage{amsfonts}
\usepackage{rotating}
\usepackage{bbold}
\usepackage{hhline}
\usepackage{braket}
\usepackage{txfonts, comment}
\usepackage{multirow}
\usepackage{appendix}
\usepackage[unicode=true,bookmarks=true,bookmarksnumbered=false,bookmarksopen=false,breaklinks=false,pdfborder={0 0 1},backref=false,colorlinks=true]{hyperref}

\hypersetup{linkcolor=magenta,urlcolor=blue,citecolor=blue,pdfstartview={FitH},hyperfootnotes=false,unicode=true}

\def\be{\begin{equation}}
\def\ee{\end{equation}}
\def\bea{\begin{eqnarray}}
\def\eea{\end{eqnarray}}

\begin{document}

\title{Random non-Hermitian Hamiltonian framework for symmetry breaking dynamics}

\author{Pei Wang}
\affiliation{Department of Physics, Zhejiang Normal University, Jinhua 321004, China}
\email{wangpei@zjnu.cn}

\begin{abstract}
We propose random non-Hermitian Hamiltonians to model the generic 
stochastic nonlinear dynamics of a quantum state in Hilbert space. 
Our approach features an underlying linearity in the dynamical equations, 
ensuring the applicability of techniques used for solving linear systems. 
Additionally, it offers the advantage of easily incorporating statistical 
symmetry, a generalization of explicit symmetry to stochastic processes. 
To demonstrate the utility of our approach, we apply it to describe real-time 
dynamics, starting from an initial symmetry-preserving state and evolving 
into a randomly distributed, symmetry-breaking final state. Our model 
serves as a quantum framework for the transition process, from 
disordered states to ordered ones, where symmetry is spontaneously broken.
\end{abstract}


\maketitle

\section{Introduction}

The Schr\"{o}dinger equation is both deterministic and linear, 
forming the foundation of quantum mechanics. Since the 1990s, however, 
efforts have been made to generalize the Schr\"{o}dinger equation into 
stochastic nonlinear differential equations~\cite{GRW,Diosi89,Bassi05,CSL,CSL2,Penrose96,Pearle99,Adler07,Adler08,
Bassi13,Pontin19,Vinante17,Donadi21,Bahrami18,Tilloy19,Gasbarri21,Vinante16,Vinante20,
Carlesso22,Zheng20,Komori20,Gisin92,Percival98,Plenio98,Daley14,Castin95,Power96,
Ates12,Hu13,Raghunandan18,Pokharel18,Weimer21,Weimer22}. There are two primary motivations 
for this generalization. The first is to provide an explanation for the objective 
wave function collapse during quantum measurement, which leads to spontaneous collapse models~\cite{GRW,Diosi89,Bassi05,CSL,CSL2,Penrose96,
Pearle99,Adler07,Adler08,Bassi13}. These models differ 
from conventional quantum mechanics and have recently seen renewed interest, 
with numerous experimental platforms being proposed to test 
them~\cite{Pontin19,Vinante17,Donadi21,Bahrami18,Tilloy19,Gasbarri21,Vinante16,Vinante20,
Carlesso22,Zheng20,Komori20}. 
The second motivation arises from the study of open quantum systems, 
where master equations for the density matrix can be unraveled into stochastic nonlinear equations~\cite{Gisin92,Percival98,Plenio98,Daley14}. These unravelled equations have 
found applications in quantum optics~\cite{Castin95,Power96,Ates12,Hu13,Raghunandan18,
Pokharel18,Weimer21,Weimer22}.
Despite these studies, a general framework for quantum stochastic nonlinear 
dynamics remains elusive, primarily due to the challenges in solving nonlinear 
equations and incorporating symmetries---such as Lorentz symmetry---into the
formalism~\cite{Myrvold17,Tumulka20,Jones20,Jones21}. 
This highlights the need for new approaches.

In this paper, we demonstrate that stochastic nonlinear dynamics can be 
equivalently generated by linear evolution operators governed by random 
non-Hermitian (RNH) Hamiltonians. While there has been extensive 
research on non-Hermitian Hamiltonians, particularly with or without 
PT-symmetry~\cite{Bender07}, in the dynamics of open quantum systems, 
including quantum scattering~\cite{Rotter09} and non-Hermitian topological 
insulators~\cite{Hatano96,Xiong18,Alvarez18,Wang18,Gong18,Kunst18,Kawabata19,Budich19,Budich20,WangZ22,Xue21,Wang24}, as well as in classical optical, mechanical, and electrical systems~\cite{Ganainy18,Kunst21}, 
these models differ from ours in that they usually do not incorporate temporal 
noise or stochastic processes, which are critical to our approach. But 
there also exists few exception~\cite{Wiersig20,Pablo24}, which appears recently.

Our method offers several advantages. It retains a hidden linearity in the 
equations, which not only simplifies solving them but also naturally reveals 
the symmetry of the system---a feature that was challenging to access 
in previous approaches. In fact, in the presence of temporal noise, the 
conventional concept of quantum symmetry (explicit symmetry) must be 
replaced by statistical symmetry~\cite{Wang22}. Statistical symmetry 
refers to the invariance of the probability distribution of an ensemble of 
quantum-state trajectories under symmetry transformations, even though 
individual random trajectories may not remain invariant.

We demonstrate the application of RNH-Hamiltonians with statistical 
symmetry by exploring real-time dynamics leading to spontaneous symmetry 
breaking (SSB). While SSB is a well-established concept in equilibrium 
statistical mechanics, its corresponding dynamical process remains poorly 
understood. Consider the $\text{Z}_2$-symmetric transverse-field Ising model 
at zero temperature. Suppose the system begins in a symmetry-preserving 
initial state, with all spins aligned in the transverse direction, corresponding 
to the ground state in the strong-field limit. As the field is withdrawn, the spins 
relax to the zero-field ground states, which are ferromagnetic states that 
spontaneously break $\text{Z}_2$ symmetry. During this evolution, the system 
must "choose" between two degenerate states (spin-up or spin-down), 
each with a 50\% probability, determined by uncontrollable environmental 
perturbations. This makes the process inherently stochastic, while still 
respecting $\text{Z}_2$ symmetry.
Furthermore, the system does not evolve into a superposition of spin-up 
and spin-down states, even though such a superposition would also possess 
the same ground-state energy due to $\text{Z}_2$ symmetry. To understand 
why the superposition is ruled out, we must account for the role of the 
environment, which selects the pointer states (spin-up or spin-down) from 
an infinite set of degenerate superposition states. To the best of our knowledge, 
no existing models explain this process.

In this paper, we show that this process can indeed be described using random RNH-Hamiltonians. To demonstrate this, we introduce an RNH term into fully connected spin models, which are known to capture symmetry-breaking physics~\cite{Sciolla11} and are mathematically tractable. These models therefore provide a useful starting point for studying RNH systems. We not only construct an exactly solvable benchmark RNH
model but also investigate more general RNH-Hamiltonians that are not strictly 
solvable. To tackle these, we develop a robust stochastic semiclassical method 
for approximating solutions. Our work lays the foundation for the RNH-Hamiltonian 
theory of stochastic nonlinear quantum dynamics.

The remainder of the paper is organized as follows. In Sec.~\ref{sec:exm}, 
we introduce the exactly solvable model to illustrate the mechanism by which the RNH term 
induces symmetry breaking during the time evolution. In Sec.~\ref{sec:stoa}, we develop 
the semiclassical approach and apply it to study an Ising-type RNH Hamiltonian, focusing 
on the role of the Hermitian component in the symmetry-breaking dynamics. Finally, 
Sec.~\ref{sec:con} summarizes our findings and outlines potential directions for future research.

\section{Exactly solvable model}
\label{sec:exm}

Our RNH models can be conveniently expressed
using an infinitesimal Hamiltonian integral~\cite{Wang22}, defined as
\be\label{eq:m:mod}
d\hat{H}_t = \hat{H}_0 \ dt + i \hat{V} \ dW_t,
\ee
where the first term represents the Hermitian Hamiltonian acting over 
an infinitesimal time interval, while the second term accounts for the 
random non-Hermitian contribution. Here, $dW_t$ denotes the differential of a Wiener process, 
and $\hat{V}$ is a Hermitian operator. The prenormalized quantum state evolves as
$\ket{ \phi_{t+dt}} = e^{-i d\hat{H}_t } \ket{ \phi_{t}}$, with the infinitesimal evolution operator
$\hat{U}_{dt} = e^{-i d\hat{H}_t}$. When $\hat{V}=0$, $\hat{U}_{dt}$ describes the 
conventional deterministic unitary evolution. However, for general $\hat{V}\neq 0$,
$\hat{U}_{dt}$ leads to a nonunitary and stochastic evolution.

In our theory, the physical state is represented by the normalized state vector,
$\ket{\psi_t}= \ket{\phi_t}/\sqrt{\braket{\phi_t|\phi_t}}$, which satisfies the following 
stochastic nonlinear equation (see Appendix~\ref{sec:app:eq} for the derivation):
\be
\begin{split}\label{eq:m:non}
& \ket{d\psi_t} = -i \hat{H}_0 dt \ket{\psi_t} + 
dW_t \left[ \hat{V} - \langle \hat{V} \rangle\right] \ket{\psi_t} \\ &
+ dt \left\{\frac{1}{2} \left[ \hat{V} - \langle \hat{V} \rangle\right]^2
- \left[ \langle\hat{V}^2 \rangle - \langle \hat{V}\rangle^2 \right] \right\}\ket{\psi_t},
\end{split}
\ee
where $\langle\hat{V} \rangle = \bra{\psi_t} \hat{V} \ket{\psi_t} $. Note the 
similarities and differences between Eq.~\eqref{eq:m:non} and the CSL model~\cite{CSL,CSL2}.
A key advantage of the RNH approach is that it avoids the need to directly solve Eq.~\eqref{eq:m:non},
Instead, we can solve the much simpler linear equation for $\ket{\phi_t}$,
and then normalize it to obtain $\ket{\psi_t}$ at the final time.

To be specific, we consider a system of $N$ spin-$1/2$ particles exhibiting $\text{Z}_2$ symmetry,
with the symmetry operator defined as $\hat{X} = \bigotimes_j \hat{\sigma}_j^x$, which flips all spins in
the $z$-direction simultaneously. Let $\hat{V} = \sqrt{\gamma} \hat{\sigma}_z$,
where $\hat{\sigma}_z=\sum_j \hat{\sigma}_j^z$ is the total spin in the $z$-direction,
and $\gamma$ represents the noise strength with the dimension of energy (or the inverse of time). 
First, we set $\hat{H}_0 = 0$, rendering the model exactly solvable.
Since $\hat{X}\hat{\sigma}_z\hat{X}=-\hat{\sigma}_z$, it is straightforward to
see that $\hat{X} d\hat{H}_t \hat{X} = -d\hat{H}_t\neq d\hat{H}_t$, meaning that
the model~\eqref{eq:m:mod} does not exhibit explicit $\text{Z}_2$ symmetry.
However, the Wiener process is symmetrically distributed around zero,
or equivalently, in probabilistic terms, $dW_t \stackrel{d}{=} -dW_t$, where
$ \stackrel{d}{=}$ denotes equality in distribution. This leads to
$\hat{X} d\hat{H}_t \hat{X} \stackrel{d}{=} d\hat{H}_t$ and, consequently,
\be
\hat{X} \hat{U}_t \hat{X} \stackrel{d}{=} \hat{U}_t,
\ee
where $\hat{U}_t$ is the evolution operator over a finite time interval. 
This defines what we term statistical symmetry.
Since the symmetry operator $\hat{X}$ is unitary (or antiunitary),
it preserves the length of vectors in Hilbert space, ensuring that the normalized 
dynamical equation~\eqref{eq:m:non} remains invariant under $\hat{X}$.
Statistical $\text{Z}_2$ symmetry implies that the ensemble of quantum-state trajectories 
retains the same distribution under the transformation $\hat{X}$.
To see this, consider a symmetry-preserving initial state $\hat{X} \ket{\psi_0} = \ket{\psi_0} $,
where all spins are aligned along the positive $x$-direction.
Under this condition, it follows that
\be
\hat{X} \ket{\psi_t} \stackrel{d}{=} \ket{\psi_t}
\ee
for arbitrary $t$.

\subsection{$N=1$ case}

We begin by analyzing the $N=1$ case of our model to understand how it 
captures the stochastic dynamics leading to symmetry-breaking states.

It is important to note that our model is not derived from the Schr\"odinger equation.
In fact, the stochastic dynamics that lead to symmetry 
breaking are expected to involve wave function collapse induced by quantum measurement 
processes. These processes are fundamentally different from the deterministic evolution 
governed by the Schr\"odinger equation, as they can produce random outcomes. 
However, there is currently no consensus on how to consistently describe continuous-time 
wave function collapse. Therefore, we leave the task of deriving our model from a more 
fundamental dynamical framework for future investigation.

The evolution operator over a finite time interval \( t \) is
$\hat{U}_t = e^{\sqrt{\gamma}\hat{\sigma}_z W_t}$. For the $N=1$ case, the system consists of a 
single spin, whose initial state is
$\ket{\psi_0} = \frac{1}{\sqrt{2}} \ket{\uparrow} + \frac{1}{\sqrt{2}} \ket{\downarrow}$.
The normalized state at time \( t \) is worked out to be
\begin{equation}\label{eq:N1psi}
\ket{\psi_t} = \frac{e^{\sqrt{\gamma} W_t} \ket{\uparrow} +
e^{-\sqrt{\gamma} W_t}\ket{\downarrow}}{\left(e^{2\sqrt{\gamma}W_t}+
e^{-2\sqrt{\gamma}W_t}\right)^{1/2}}.
\end{equation}
Equation~\eqref{eq:N1psi} describes a random state vector on the Bloch sphere. 
To understand its asymptotic behavior as \( t \to \infty \), we consider the statistical 
properties of the Wiener process \( W_t \), which follows a Gaussian distribution 
with zero mean and variance \( t \).
As \( t \to \infty \), the probability that \( |W_t| < \Omega \) for arbitrarily big \( \Omega \) vanishes, 
while the probabilities for \( W_t > \Omega \) and \( W_t < -\Omega \) both approach \( 1/2 \). 
Simply speaking, we have $W_t \to \pm \infty$ in the limit $t\to\infty$.
In the case where \( W_t \to \infty \), the state~\eqref{eq:N1psi} approaches \( \ket{\uparrow} \); 
conversely, when \( W_t \to -\infty \), it approaches \( \ket{\downarrow} \). Therefore, 
in the long-time limit, the system collapses into either \( \ket{\uparrow} \) or \( \ket{\downarrow} \) 
with equal probability.
This result captures the essence of symmetry-breaking dynamics in a two-level system, 
where the superposition between \( \ket{\uparrow} \) and \( \ket{\downarrow} \) is entirely lost. 
It thus supports the interpretation of our model as describing symmetry breaking.

\subsection{Symmetry-breaking dynamics at finite $N$}

\begin{figure}[htp]
\centering
\vspace{0.1cm}
\includegraphics[width=.45\textwidth]{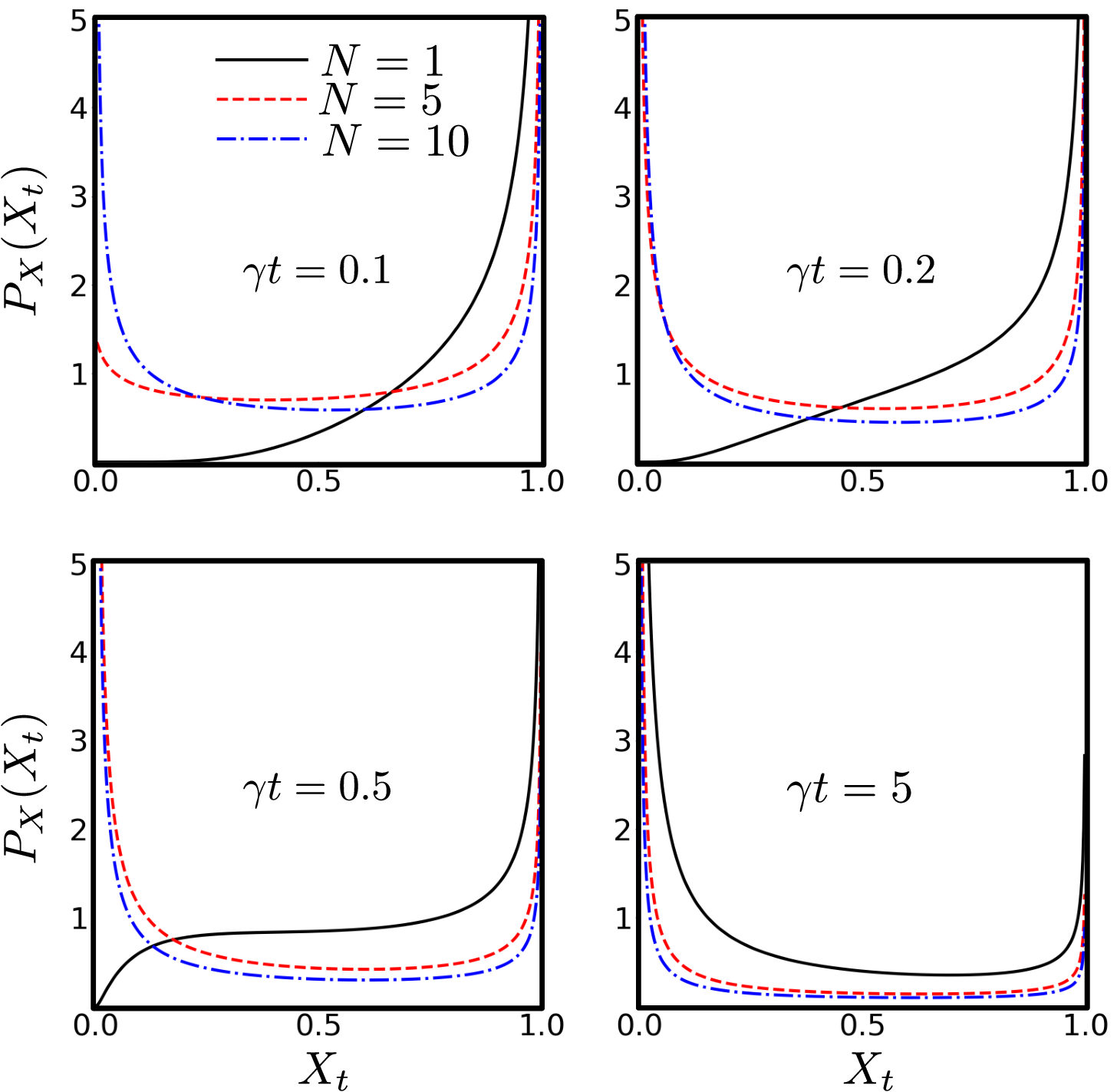}
\caption{Probability density of $X_t$ at various time points, plotted in separate panels. 
Each line style and color represents a different number of spins $N$, as indicated in the legend.}
\label{fig:XtP}
\end{figure}

Let us now study the dynamics of the system for arbitrary finite $N$. The normalized quantum state at time $t$ is given by
\begin{equation}\label{eq:fN:pt}
\ket{\psi_t} = \bigotimes_{j=1}^N \frac{e^{\sqrt{\gamma} W_t} \ket{\uparrow}_j +
e^{-\sqrt{\gamma} W_t} \ket{\downarrow}_j}{\left(e^{2\sqrt{\gamma}W_t} +
e^{-2\sqrt{\gamma}W_t}\right)^{1/2}}.
\end{equation}
We observe that $\ket{\psi_t}$ remains a product state, where each spin is in the same state. According to the previous analysis, in the limit $t \to \infty$, the system evolves into either the fully polarized state $\prod_j \ket{\uparrow}_j$ or $\prod_j \ket{\downarrow}_j$, each with probability $50\%$.

The superposition between the spin-up and spin-down states is gradually lost as time progresses. To quantify this, we compute the expectation value of the operator
$\hat{X} = \bigotimes_{j=1}^N \hat{\sigma}^x_j$,
which we denote by $X_t$. This observable reflects the degree of quantum coherence between the up and down spin components. When $X_t = 1$, the coherence is fully preserved; when $X_t = 0$, it is completely lost. A straightforward calculation yields
\begin{equation}\label{eq:Xxx}
X_t = \langle \psi_t | \hat{X} | \psi_t \rangle = \left[ \cosh \left(2\sqrt{\gamma} W_t \right) \right]^{-N}.
\end{equation}
Since $W_t \to \pm\infty$ as $t \to \infty$, we have $X_t \to 0$ due to the exponential growth of the $\cosh$ function in Eq.~\eqref{eq:Xxx}.

We further compute the probability density function of $X_t$, which takes the form:
\begin{equation}
\begin{split}
P_X(X_t = x) = \ & \frac{1}{N x \sqrt{2\pi \gamma t \left(1 - x^{\frac{2}{N}}\right)}} \\
& \times \exp\left\{
-\frac{1}{8\gamma t} \left[\ln \left(x^{-\frac{1}{N}} + \sqrt{x^{-\frac{2}{N}} - 1}\right)\right]^2
\right\},
\end{split}
\end{equation}
valid for $ x \in [0,1] $.

In Fig.~\ref{fig:XtP}, we plot the probability density of $X_t$ at different times for various values of $N$, namely $N=1$, $5$, and $10$. At the initial time $t=0$, we have $X_0 = 1$ with probability $100\%$. As time increases, the distribution of $X_t$ gradually shifts from $x=1$ toward $x=0$, indicating a loss of coherence. The rate at which $X_t$ decays to zero increases with the number of spins $N$. For $N=1$ (black solid line), the transition is relatively slow, while for $N=10$ (blue dash-dotted line), the probability concentrates near zero much more rapidly.

For any finite $N$, the wavefunction asymptotically collapses into either the all-spin-up or all-spin-down state as $t \to \infty$. This is physically reasonable. Notably, our model describes symmetry-breaking dynamics at both zero external field and zero temperature. Consider, for instance, the Ising model, a prototypical system exhibiting spontaneous $Z_2$ symmetry breaking. At zero field and zero temperature, the ground state is doubly degenerate: either all spins up or all spins down, each occurring with equal probability. This holds not only in the thermodynamic limit ($N \to \infty$) but also for any finite $N$.

At zero temperature, the system relaxes to one of these ground states regardless of the system size, in agreement with the long-time behavior predicted by our model. Thus, the stochastic dynamics described here capture the essential features of symmetry breaking even at finite $N$.

\subsection{Symmetry-breaking dynamics as $N\to\infty$}

\begin{figure}[htp]
\centering
\vspace{0.1cm}
\includegraphics[width=.49\textwidth]{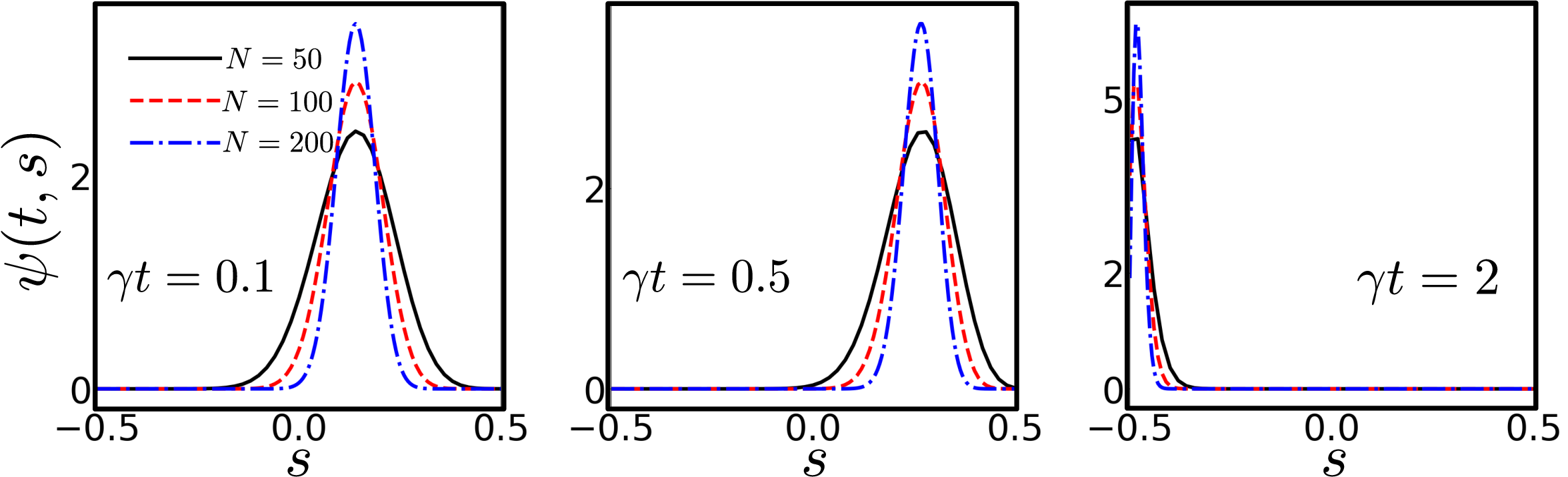}
\caption{Time evolution of the normalized wave function from 
$\gamma t = 0.1$ to $\gamma t = 2$, displayed from left to right. Comparisons for different 
values of $N$ are shown using distinct line styles and colors.}
\label{fig:psN}
\end{figure}

From the above analysis at finite $N$, we see that the model~\eqref{eq:m:mod} 
with $\hat{V} = \sqrt{\gamma} \hat{\sigma}_z$ describes a dynamical process leading to SSB. To gain deeper insight into this process, we note that the thermodynamic limit $N \to \infty$ plays a crucial role in the study of SSB. We therefore proceed to analyze the quantum state evolution in this limit.
Before taking the limit $N \to \infty$, we transform to the Dicke basis $\ket{s}$, where 
$s = \sum_j \sigma^z_j / (2N)$ represents the average magnetization. In this basis, the wave packet is given by
$\phi(t, s) = \braket{s | \phi_t} = e^{2Ns \sqrt{\gamma} W_t} \psi_0(s)$,
where the initial wave function is expressed as
\be\label{eq:mt:p0s}
\psi_0(s) \propto \sqrt{\frac{N!}{\left[N\left(\frac{1}{2}+s\right)\right]!
\left[N\left(\frac{1}{2}-s\right)\right]!}}.
\ee
Further details on the solution of the model in the Dicke basis are provided in Appendix~\ref{sec:app:st}. We also note that the symmetry transformation
\be
\hat{X} \ket{s} = \ket{-s} 
\ee
holds in this representation.

For sufficiently large $N$, applying Stirling's approximation, we find that the normalized wave function $\psi(t, s)$ develops a sharp peak with a width scaling as $\sim 1/\sqrt{N}$, which collapses to a delta function in the limit $N \to \infty$. In other words, the wave packet becomes increasingly localized at a specific point in $s$-space as $N$ grows. In Fig.~\ref{fig:psN}, we plot the wave function for different values of $N$, as time evolves from $\gamma t = 0.1$ to $\gamma t = 2$. It is evident that the width of the wave packet decreases rapidly with increasing $N$. This behavior is consistently observed at both early and late times, providing strong support for our use of Stirling's approximation.

\begin{figure}[htp]
\centering
\vspace{0.1cm}
\includegraphics[width=.49\textwidth]{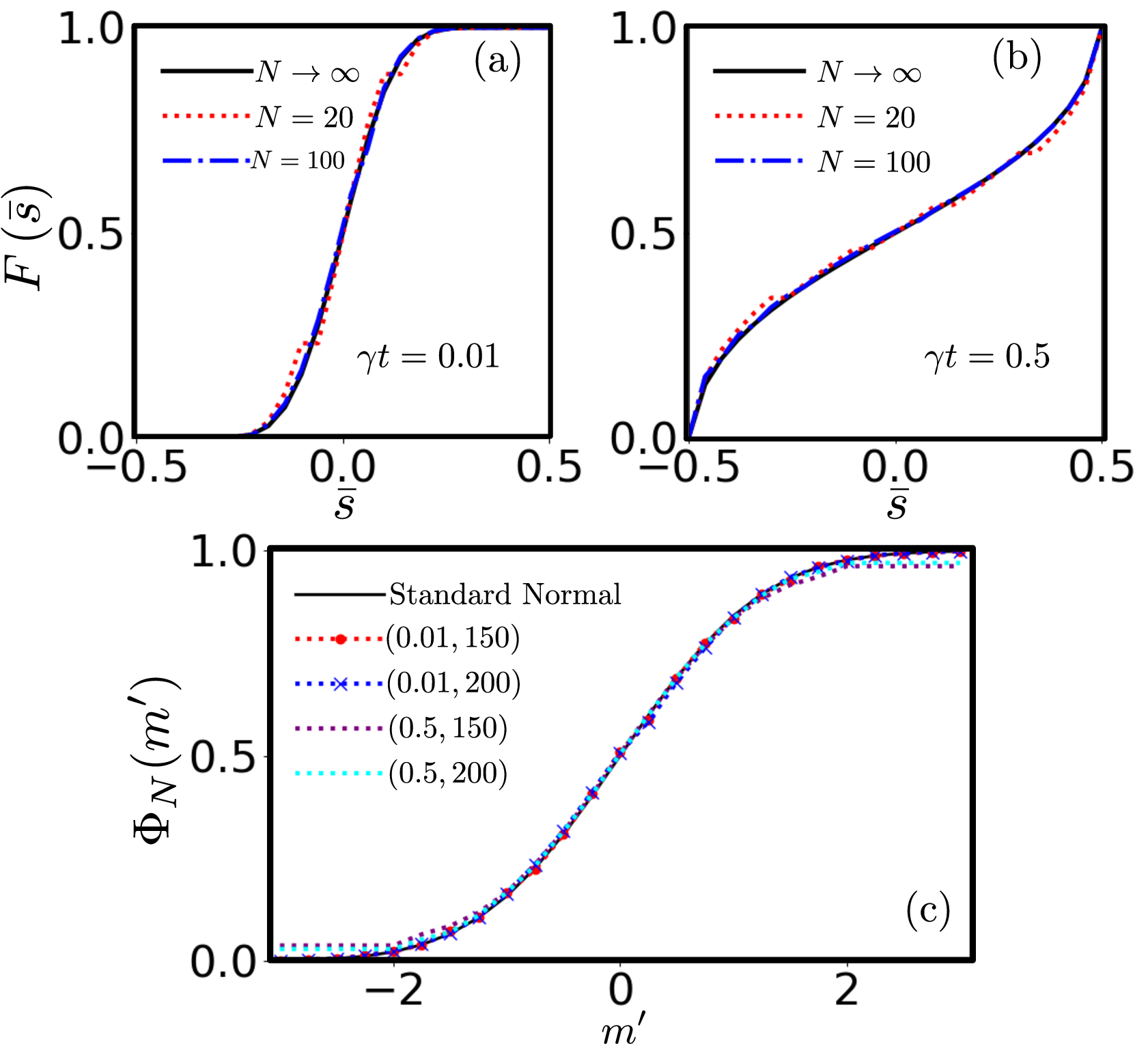}
\caption{(a,b) Cumulant distribution of $\bar{s}$ for finite $N$ and in the thermodynamic limit $N \to \infty$, distinguished by different line styles and colors. Panels (a) and (b) correspond to $\gamma t = 0.01$ and $\gamma t = 0.5$, respectively. (c) The rescaled cumulant distribution function $\Phi_N(m')$ is plotted for various pairs of $\left(\gamma t, N \right)$, represented by dotted lines with different colors and markers. In all cases, $\Phi_N(m')$ collapses onto the cumulant function of a standard normal distribution, shown as the black solid line.}
\label{fig:fNPd}
\end{figure}

Due to the stochastic nature of the dynamics, the peak location of the wave packet 
becomes a time-dependent random variable, $\bar{s}$, defined
by $\partial_s \left| \psi(t,s)\right|^2 = 0$ at ${s=\bar{s}}$. As $N\to\infty$, a straightforward 
calculation gives (see Appendix~\ref{sec:app:st} for the derivation)
\be\label{eq:bars:ex}
\bar{s} = \frac{1}{2} \tanh \left( \tilde{W}_t\right),
\ee
where $\tilde{W}_t \equiv 2 \sqrt{\gamma} W_t$. The probability distribution of $\bar{s}$ can be more
conveniently expressed using the rescaled magnetization
$\bar{m} =  \ln \frac{1+2\bar{s}}{1-2\bar{s}}$,
a one-to-one mapping from $\bar{s}\in\left(-1/2,1/2\right)$ to $\bar{m} \in (-\infty,\infty)$.
The variable $\bar{m}$ follows a Gaussian distribution with mean zero and variance 
$16\gamma t$. The cumulant distribution of $\bar{s}$ is
\be\label{eq:inFF}
F(\bar{s}) = \Phi \left( \bar{m}\left(\bar{s}\right)/ \sqrt{16\gamma t}\right), 
\ee
where $\Phi$ is the cumulative distribution function of the normal Gaussian distribution.

On the other hand, using Eq.~\eqref{eq:fN:pt}, we can numerically simulate the wave function $\psi(t,s)$
for finite $N$, and determine the location of the maximum of $\left| \psi(t,s) \right|^2$
along the $s$-axis, which is discretized at finite $N$. By performing $5\times 10^4$ independent samples,
we obtain the cumulant distribution at finite $N$, as shown in Fig.~\ref{fig:fNPd}(a) and 
Fig.~\ref{fig:fNPd}(b). For comparison,
we also plot the cumulant distribution in the thermodynamic limit $N \to \infty$, given by Eq.~\eqref{eq:inFF}.
It is evident that as $N$ increases, the finite-$N$ cumulant distribution gradually approaches the
analytic form given by Eq.~\eqref{eq:inFF}. Specifically, the $N = 100$ curve (blue dash-dotted line)
aligns more closely with the $N \to \infty$ result than the $N = 20$ curve (red dotted line). Furthermore,
the difference between the $N = 100$ and $N \to \infty$ curves becomes negligible at all times shown
(the left and right panels correspond to early time $\gamma t = 0.01$ and later time $\gamma t = 0.5$,
respectively). 
An alternative and more effective way to verify the validity of Eq.~\eqref{eq:inFF} is by plotting the rescaled cumulant function $\Phi_N$, defined as  
$\Phi_N(m') \equiv F(\bar{s})$, with $\bar{s} = \frac{1}{2} \cdot \frac{e^{m'\sqrt{16\gamma t}} - 1}{e^{m'\sqrt{16\gamma t}} + 1}$. 
As shown in Fig.~\ref{fig:fNPd}(c), $\Phi_N(m')$ for different values of $t$ (represented by dotted lines in various colors) collapses onto the cumulant distribution function of the standard normal distribution (black solid line) in the limit $N \to \infty$. These results confirm that our analytical expression in Eq.~\eqref{eq:inFF} accurately captures the cumulant distribution in the large-$N$ limit.

Note that the probability density $P(\bar{s}) = dF(\bar{s})/d\bar{s}$, which is the derivative of the cumulant distribution, provides a more direct representation of the distribution of $\bar{s}$. However, $P(\bar{s})$ is more challenging to obtain numerically compared to $F(\bar{s})$, as it requires a larger number of samples to achieve a smooth and reliable result. In contrast, the cumulant distribution $F(\bar{s})$ converges more quickly with fewer samples. Therefore, we have used the cumulant distribution for comparing our finite-$N$ numerical results with the analytical expression in Fig.~\ref{fig:fNPd}.

\begin{figure}[htp]
\centering
\vspace{0.1cm}
\includegraphics[width=.47\textwidth]{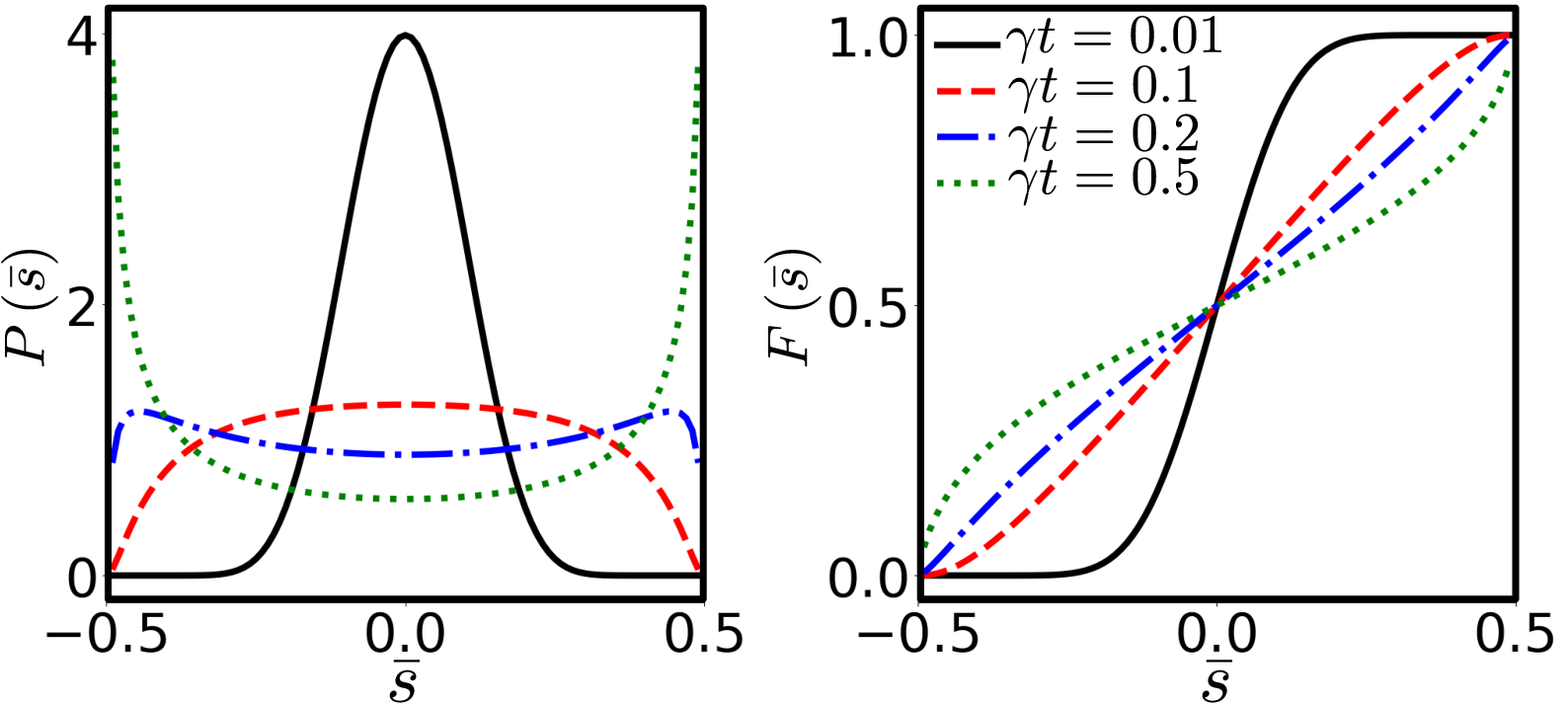}
\caption{Probability density (left panel) and cumulant distribution functions (right panel) at different times
for $N\to\infty$.}
\label{fig:strict}
\end{figure}

Figure~\ref{fig:strict} illustrates both the probability density $P(\bar{s})$ (see Appendix~\ref{sec:app:st} for its expression) and the cumulant distribution $F(\bar{s})$ in the thermodynamic limit, comparing the results at different evolution times. At time $t=0$, the probability density is $P(\bar{s})  = \delta(\bar{s})$, corresponding to
zero magnetization, reflecting the explicit $\text{Z}_2$ symmetry of the initial state. 
As time progresses, $P(\bar{s})$ flattens (see Fig.~\ref{fig:strict} the left panel),
but remains symmetric about zero due to statistical symmetry, which enforces
$P(\bar{s}) = P(-\bar{s})$. Over time, the probability shifts from $\bar{s}=0$
to $\bar{s}=\pm 1/2$, eventually forming two peaks at $\bar{s}=\pm 1/2$.
In the limit as $t\to\infty$, the probability distribution becomes $P(\bar{s})  
= \frac{1}{2} (\delta(\bar{s}+\frac{1}{2}) +\delta(\bar{s}-\frac{1}{2})$. 
The all-spin-up and all-spin-down states both occur with probability $1/2$,
but superpositions of these states do not persist, as desired.

While the cumulant distribution has a one-to-one correspondence with the probability density (see the right panel of Fig.~\ref{fig:strict}), it provides an alternative view of the probability concentration. At $t=0$, $F(\bar{s})$ is a step function with a jump at $\bar{s}=0$, reflecting that all the probability is concentrated at $\bar{s}=0$. As time progresses, the slope of $F$ around $\bar{s}=0$ gradually decreases, while the slope near $\bar{s} = \pm 1/2$ increases, indicating a migration of probability from $\bar{s}=0$ to $\bar{s} = \pm 1/2$. In the limit $t \to \infty$, $F(\bar{s})$ becomes a piecewise constant function, showing two sharp jumps at $\bar{s} = \pm 1/2$, corresponding to the final symmetry-breaking states.

Notably, the states $\ket{\bar{s}=\pm 1/2}$ are ferromagnetic and
spontaneously break $\text{Z}_2$ symmetry, yet the overall distribution 
respects statistical symmetry, as the probabilities of $\bar{s}=1/2$ and $\bar{s}=-1/2$ are equal. 
The timescale for symmetry breaking is characterized by $1/\gamma$.
For $t\ll 1/\gamma$, the probability remains concentrated around $\bar{s}=0$
(the symmetry-preserving state), while for $t\gg 1/\gamma$, it converges at
$\bar{s}=\pm 1/2$ (the SSB states).

A few comments are in order. First, we emphasize that in our model, the limits $N \to \infty$ and $t \to \infty$ are commutative. As discussed above, regardless of whether the thermodynamic limit ($N \to \infty$) or the long-time limit ($t \to \infty$) is taken first, the wave function ultimately relaxes to either the all-spin-up or all-spin-down state. This commutativity arises because our model describes a system deep within the symmetry-broken phase---specifically, at zero external field and zero temperature. In this regime, symmetry breaking is robust, and the dynamics naturally lead to one of the ordered states. 
However, we note that our model is not applicable in the vicinity of the critical point. Near the phase transition, critical slowing down becomes significant, and the two limits ($N \to \infty$ and $t \to \infty$) may no longer commute. In such cases, the system’s relaxation dynamics can become sensitive to the order in which these limits are taken.

Second, our specific choice of $\hat{V} = \sqrt{\gamma} \sum_j \hat{\sigma}_j^z$ in the RNH term is responsible for the survival of only the all-spin-up and all-spin-down states. These two configurations correspond to the eigenstates of $\hat{V}$ with the maximum and minimum eigenvalues, respectively. 
If, instead, the RNH term is modified so that each spin experiences independent decoherence---for example, by allowing $dW_t$ to be site-dependent. In that case, the wave function could collapse into arbitrary spin configurations, with each spin independently choosing either the up or down state. However, superpositions of different spin configurations would still be suppressed due to decoherence.
Such a situation effectively corresponds to a high-temperature mixed state and falls outside the low-temperature, symmetry-breaking regime described by our current model. The investigation of this alternative scenario, along with other possible forms of RNH Hamiltonians, is left for future work.

Finally, we point out that the probability density at zero magnetization, $P(\bar{s}=0)$, decays as a power law in time, following a $1/\sqrt{t}$ scaling, rather than an exponential decay. This slow relaxation originates from the fact that $\bar{s}=0$ corresponds to $W_t = 0$, i.e., the Brownian motion remaining at the origin. Since $W_t$ follows a Gaussian distribution with variance proportional to $t$, the probability density at $W_t = 0$ scales as $1/\sqrt{t}$.
The power-law behavior of magnetization is a consequence of the coupling between the nonHermitian operator and $W_t$. This suggests that separating the nonHermitian Hamiltonian from the random noise may lead to different decay behaviors. Exploring this distinction could provide valuable insight and represents a promising direction for future investigation.

\subsection{Fluctuation time between spin-up and spin-down states}

We now investigate the fluctuation time between up and down magnetization states. From the preceding analysis, we know that the system eventually settles into one of the two symmetry-broken states: either all spins up or all spins down. Suppose the system is initially in a state where nearly all spins are down. Although rare, there remains a possibility that, due to stochastic fluctuations, the system may eventually transition to a state where almost all spins are up. Since spin dynamics in our model are governed by stochastic processes, the time required for such a fluctuation is a random variable. To study this, we use Eq.~\eqref{eq:bars:ex}, where $\bar{s}$ quantifies the degree of spin alignment: as $\bar{s} \to \pm 1/2$, all spins are aligned in the up or down direction, respectively.

Equation~\eqref{eq:bars:ex} describes the stochastic evolution of $\bar{s}$. Suppose that at some time $t_a$, the magnetization reaches $\bar{s}(t_a) = -\mathcal{M}$, where $\mathcal{M}$ is close to $1/2$. We define the fluctuation time $T$ as the time it takes for $\bar{s}(t)$ to first reach $+\mathcal{M}$, i.e., $\bar{s}(t_a + T) = +\mathcal{M}$. Since the trajectory of $\bar{s}(t)$ is stochastic, it may hit $+\mathcal{M}$ multiple times during its evolution. Our key idea is to define the fluctuation time as the first hitting time of $\bar{s}(t)$ reaching $+\mathcal{M}$ after being at $-\mathcal{M}$.

Solving Eq.~\eqref{eq:bars:ex}, we find $W_{t_a} = -\frac{1}{2\sqrt{\gamma}} 
\tanh^{-1}\left(2\mathcal{M}\right)$ and $W_{t_a+T} = \frac{1}{2\sqrt{\gamma}} \tanh^{-1}\left(2\mathcal{M}\right)$.
Using the property of independent increments of the Wiener process, we have
\be
W_T = W_{t_a+T} - W_{t_a} = \frac{\tanh^{-1}\left(2\mathcal{M}\right)}{\sqrt{\gamma}}.
\ee
Thus, the fluctuation time $T$ corresponds to the first hitting time of a standard Brownian motion reaching the point $\tanh^{-1}\left(2\mathcal{M}\right)/\sqrt{\gamma}$.
This classic problem has a well-known solution. The probability density of the first hitting time $T$ is given by
\be
P_T (T) = \frac{\tanh^{-1}\left(2\mathcal{M}\right)}{\sqrt{2\pi \gamma T^3}} 
\exp\left\{-\frac{\left[\tanh^{-1}\left(2\mathcal{M}\right)\right]^2}{2\gamma T}\right\}.
\ee
Note that the mean of $T$ does not exist, as the integral diverges. However, the maximum of this probability density occurs at
\be
T_{\mathrm{max}} = \frac{\left[\tanh^{-1}\left(2\mathcal{M}\right)\right]^2}{3\gamma},
\ee
which we identify as the typical fluctuation time. As $\mathcal{M} \to 1/2$, $T_{\mathrm{max}}$ diverges. This is physically intuitive: the all-spin-up and all-spin-down states are eigenstates of the operator $\hat{V}$, and once the system reaches either of these states, it becomes trapped and cannot transition away.

We emphasize again that this analysis applies to symmetry-breaking dynamics deep within the ordered phase, far from the critical point. Therefore, we do not expect to observe critical fluctuations or fluctuation times that grow with $N$.

\section{Stochastic semiclassical approach}
\label{sec:stoa}

Next, we examine the impact of a Hermitian term,
$\hat{H}_0$, on the dynamics. Specifically, we consider the prototypical transverse-field Ising Hamiltonian
with all-to-all couplings:
\be
\hat{H}_0 = -\frac{J}{N} \hat{\sigma}_z^2 + h \hat{\sigma}_x, 
\ee
where $J>0$ represents the ferromagnetic coupling and $h>0$ represents the transverse field.
Here, $\hat{\sigma}_z = \sum_j \hat{\sigma}^z_j$ and $\hat{\sigma}_x = \sum_j \hat{\sigma}^x_j$ are the sums of spin operators over all sites. The term $\hat{\sigma}_z^2 = \sum_{i,j} \hat{\sigma}^z_i \hat{\sigma}^z_j$ thus represents pairwise couplings between all spins. The coupling strength $-J/N$ is uniform, meaning it is the same for every pair of spins, regardless of which two spins are involved.
It is easy to see that $\hat{H}_0= \hat{X} \hat{H}_0\hat{X} $, which displays $\text{Z}_2$ symmetry. 
Therefore, adding it to $d\hat{H}_t$ preserves the statistical $\text{Z}_2$-symmetry.
Furthermore, the Ising model with all-to-all couplings is exactly solvable and exhibits strict mean-field behavior. This makes it an ideal starting point for studying our RNH dynamics, before moving on to more realistic, but technically challenging, models with nearest-neighbor couplings. The ground state of the Hamiltonian $\hat{H}_0$ undergoes a quantum phase transition at $h = 2J$, where the $\text{Z}_2$ symmetry is spontaneously broken for $h < 2J$.

When $h\neq 0$, Eq~\eqref{eq:m:mod} is no longer strictly solvable. To obtain an approximate 
solution for large $N$, we develop the stochastic semiclassical approach.
The key idea is that, for sufficiently large $N$, the wave function at any time $t$ exhibits a sharp peak
in $s$-space. Consequently, we can focus on the location of the peak, denoted by $\bar{s}$,
while neglecting the detailed form of $\phi(t,s)$. To determine $\bar{s}(t)$, we first express $\phi(t,s)$ as
$\phi(t,s) = \exp \left\{ -N \left( g(t,s) - i\theta(t,s) \right) \right\}$, 
and derive the dynamical equations for $g(t,s)$ and $\theta(t,s)$. The peak location $\bar{s}(t)$
corresponds to the minimum of $g(t,s)$. By expanding $g(t,s)$ and $\theta(t,s)$
around $\bar{s}$ using a Taylor series and substituting into the dynamical equations, we find
(see Appendix~\ref{sec:app:ssa} for the detailed derivation):
\be\label{eq:m:dsdp}
\begin{split}
& d\bar{s} = d\left(\frac{1}{2}\tanh\left(\tilde{W}_t\right)\right)
-\tilde{f}(\bar{s})\sin\left(\bar{p}\right) \ dt, \\ &
d\bar{p} = 4h^2 \ \bar{s} \tilde{f}(\bar{s})^{-1} \cos(\bar{p})  \ dt
+ 4Jt \ d\left( \tanh\left(\tilde{W}_t\right)\right) + 8J\bar{s} \ dt,
\end{split}
\ee
where $\tilde{f}(s)=2h \sqrt{\frac{1}{4}-s^2}$, and $\bar{p} = \partial_s \theta\left|_{s=\bar{s}} \right.$
represents the classical momentum.
As $\gamma=0$, Eq.~\eqref{eq:m:dsdp} reduces to the classical equations of motion
governed by the Hamiltonian $H_0 (\bar{s},\bar{p})= \tilde{f}(s) \cos \bar{p} - 4J\bar{s}^2$,
which is well known in the semiclassical treatment of fully-connected models~\cite{Sciolla11}.
A systematic perturbative approach can be applied to solve Eq.~\eqref{eq:m:dsdp}. 
To first order of $h$, the solution is given by:
\be\label{eq:m:sol}
\begin{split}
& \bar{s}(t) \approx \bar{s}_0(t) -h \int^t_0
d\tau \sqrt{X_\tau} \sin
\left[ 8J\tau \bar{s}_0(\tau) \right], \\
& \bar{p}(t) \approx 8J  t \bar{s}_0(t) + 4h \int^t_0 d\tau \frac{\bar{s}_0(\tau) }
{\sqrt{ X_\tau}}  \cos \left[ 8J \tau \bar{s}_0(\tau) \right],
\end{split}
\ee
where $\bar{s}_0(t) = \frac{1}{2}\tanh\left(\tilde{W}_t\right)$ is the exact
solution for $h=0$, and $X_\tau = 1-4\bar{s}_0(\tau)^2$. Our numerical simulations
verify that solutions~\eqref{eq:m:sol} provide good approximations (see Appendix~\ref{sec:app:ssa}), 
except when $\gamma t\gg 1$, at which point they predict unphysical results ($\left|\bar{s} \right| > 1/2$).
The behavior for $\gamma t\gg 1$ will be discussed separately.
We emphasize that our approach is applicable to general RNH-Hamiltonians in a broad 
class of full-connected models.

\begin{figure}[htp]
\centering
\vspace{0.1cm}
\includegraphics[width=.48\textwidth]{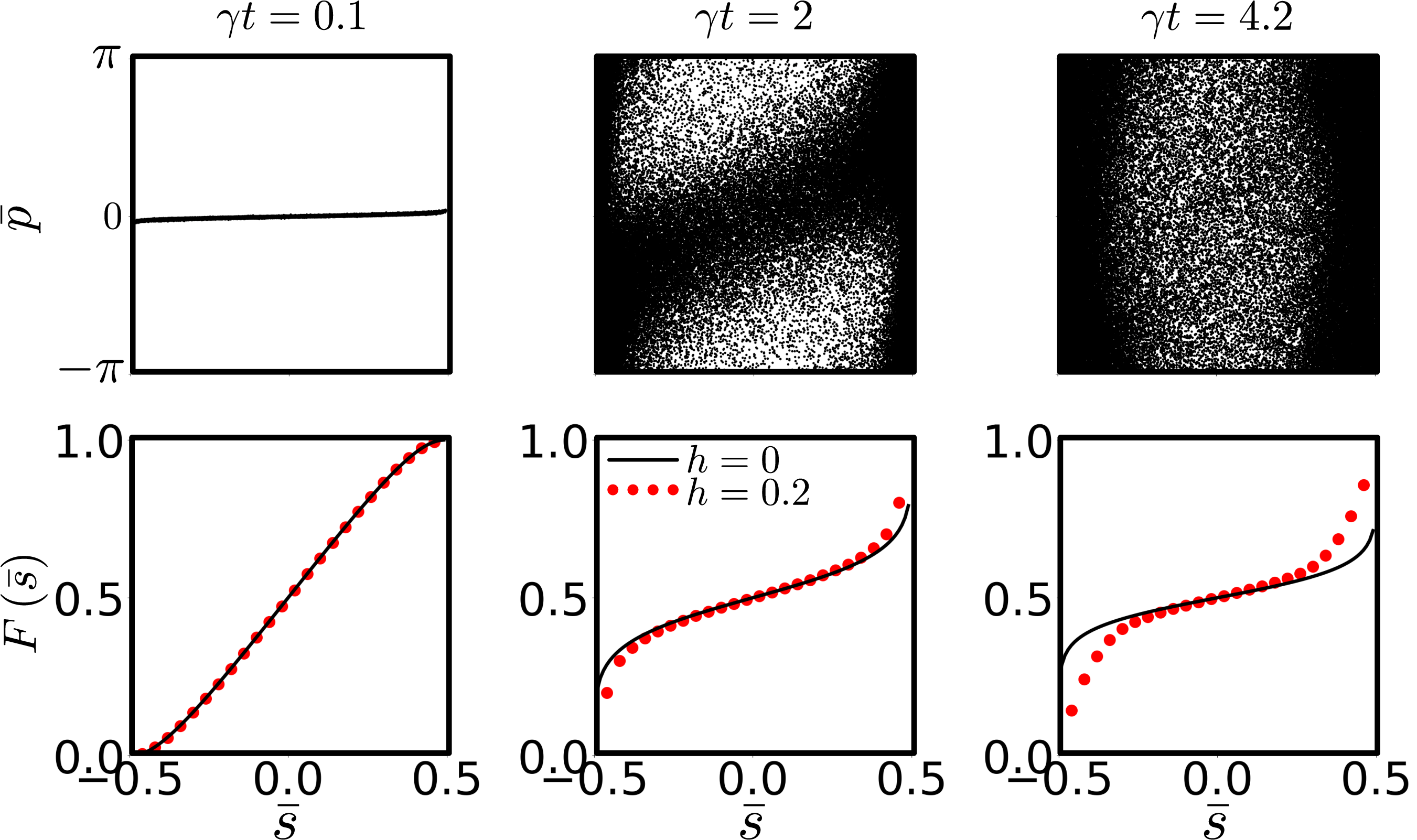}
\caption{Top panels: Samples of random variables $\left(\bar{s},\bar{p}\right)$ plotted in the $\bar{s}-\bar{p}$
plane at different times---$\gamma t=0.1$, $2$, and $4.2$ from left to right, obtained by using Eq.~\eqref{eq:m:sol}. Parameters are chosen as $h=0.2$ and $J=0.1$. $\left(\bar{s},\bar{p}\right)$ are sampled $10^5$ times.
Bottom panels: The corresponding cumulant distribution functions (red dots). For comparison, the black
solid line represents the cumulant distribution function at $h=0$. }
\label{fig:sca}
\end{figure}

\subsection{Real-time dynamics when $h\neq 0$}

Next, we set $\gamma = 1$ as the unit of energy and examine how the parameters $h$ and $J$ influence the probability migration from the symmetry-preserving state to the SSB states, using Eq.~\eqref{eq:m:sol}.

The probability migration is reflected in the slope of the cumulant distribution $F(\bar{s})$, or equivalently, in the probability density $P(\bar{s})$, at $\bar{s} = \pm 1/2$, which increases from zero to infinity as time progresses. Alternatively, it can be seen in the slope at $\bar{s} = 0$, which decreases from infinity toward zero. Note that the case of $h = 0$ is exactly equivalent to the case where $\hat{H}_0 = 0$, which has already been discussed earlier. In this situation, the Dicke basis states $\ket{s}$ are eigenstates of $\hat{H}_0$, so the Hermitian Hamiltonian contributes only a phase to the wave function $\psi(t,s)$ and therefore does not affect the location of the maximum of $\left| \psi(t,s) \right|^2$, i.e., the value of $\bar{s}$ (see Appendix~\ref{sec:app:seh} for the detail).

Figure~\ref{fig:sca} (bottom panels) shows $F(\bar{s})$ for $h = 0.2$ (red dots) at different times. For comparison, the exact results for $h = 0$ (black solid lines, identical to those shown in Fig.~\ref{fig:strict}) are also plotted, which exhibit perfect probability migration, as previously discussed.
We observe that the results for $h = 0.2$ closely follow the exact $h = 0$ solution up to approximately $t \sim 2$. Beyond this point, noticeable deviations begin to appear, becoming more pronounced at later times (e.g., $t = 4.2$). This growing deviation indicates that the probability migration is progressively hindered due to the presence of a nonzero transverse field $h$.
At $ t =4.2$, the red dots show a steeper slope at $\bar{s}=0$ and a shallower one
at $\bar{s}=\pm 1/2$ compared to the black solid line,
suggesting that the development of SSB is being impeded by the transverse field $h$.
This aligns with the fact that quantum fluctuations induced by the transverse field disrupt
the ferromagnetic order in equilibrium. We explored different parameter values (see Appendix~\ref{sec:app:rts})
and found qualitatively similar behavior, with the deviation occuring earlier (later) as $h$ increases (decreases).

Although the symmetry-breaking information is primarily encoded in the distribution of $\bar{s}$, the dynamical equations governing $\bar{s}$ and $\bar{p}$ are coupled, as described by Eq.~\eqref{eq:m:dsdp}. To gain deeper insight into the dynamical behavior of $\bar{s}$---especially at later times when Eq.~\eqref{eq:m:sol} no longer holds---it is helpful to examine the stochastic dynamics of the pair $\left(\bar{s},\bar{p}\right)$.

Figure~\ref{fig:sca} (top panels) presents the distribution of $\left(\bar{s},\bar{p}\right)$ in phase space, based on $10^5$ samples. The results reveal a clear correlation between $\bar{s}$ and $\bar{p}$ at early and intermediate times (e.g., $\gamma t =0.1, 2$), which gradually diminishes as time progresses. This early-time correlation can be understood using Eq.~\eqref{eq:m:dsdp}. At early times, the $\gamma$-term (i.e., the terms involving $\tilde{W}_t$) dominates, pushing $\bar{s}$ away from the origin and stretching the point cloud in the $\bar{p}$-direction in a counterclockwise manner (as seen from the expression for $\bar{p}$ in Eq.~\eqref{eq:m:sol}). Meanwhile, the Hermitian Hamiltonian $\hat{H}_0$ causes a counterclockwise rotation of the $\left(\bar{s},\bar{p}\right)$ points, consistent with classical Hamiltonian dynamics governed by $H_0(\bar{s}, \bar{p})$ (see Appendix~\ref{sec:app:ssds} for detailed analysis and figures).
The combined effects of the $\gamma$-term and $\hat{H}_0$ lead to an asymmetric distribution at intermediate times such as $\gamma t = 2$. However, at sufficiently long times ($\gamma t = 4.2$), both the asymmetry and the correlation between $\bar{s}$ and $\bar{p}$ disappear. The resulting distribution becomes nearly uniform along the $\bar{p}$-direction, indicating that $P(\bar{s}, \bar{p})$ becomes effectively independent of $\bar{p}$ in the long-time limit. Such an independence is a crucial property that facilitates the analysis of the long-time behavior of $\bar{s}$, as discussed below.

Our perturbative results~\eqref{eq:m:sol} break down when $t\gg 1/h$. To gain a 
qualitative understanding of the steady-state distribution, we revisit Eq.~\eqref{eq:m:dsdp}.
In the limit $t\to\infty$, $\tanh\left(\tilde{W}_t\right) \to \pm 1$ and $d\tanh\left(\tilde{W}_t\right) \to 0$,
allowing us to neglect the $\gamma$-terms. In this case, Eq.~\eqref{eq:m:dsdp} simplifies 
to the classical dynamical equation governed by $H_0(\bar{s},\bar{p})$, and the corresponding 
Fokker-Planck equation can be easily worked out (see Appendix~\ref{sec:app:ssds} for the detail). 
We use $P(t,\bar{s},\bar{p})$ to denote the probability
density in the phase space. By setting $\partial_t P =0$, we find the condition 
for the steady-state distribution.
Based on our earlier observation---$\partial_{\bar{p}} P(\bar{s},\bar{p}) =0$ in the steady-state 
limit---we conclude that $\partial_{\bar{s}} P(\bar{s},\bar{p}) = 0$ as long as $\bar{s}\neq 1/2$. This
implies that the steady-state distribution is surprisingly simple: the probability
density for $-1/2<\bar{s}<1/2$ must be constant, denoted as $\Delta$, with $0\leq \Delta\leq 1$.
The probabilities of $\bar{s}=\pm 1/2$ are both $\left(1-\Delta\right)/2$ due to statistical symmetry.
The parameter $\Delta$ is called the residue probability, representing the portion of the probability
that has not migrated to the SSB state $\ket{\bar{s}=\pm 1/2}$. The exact solution for $h=0$
corresponds to $\Delta =0$, indicating no residue probability and complete symmetry breaking.

\begin{figure}[htp]
\centering
\vspace{0.1cm}
\includegraphics[width=.48\textwidth]{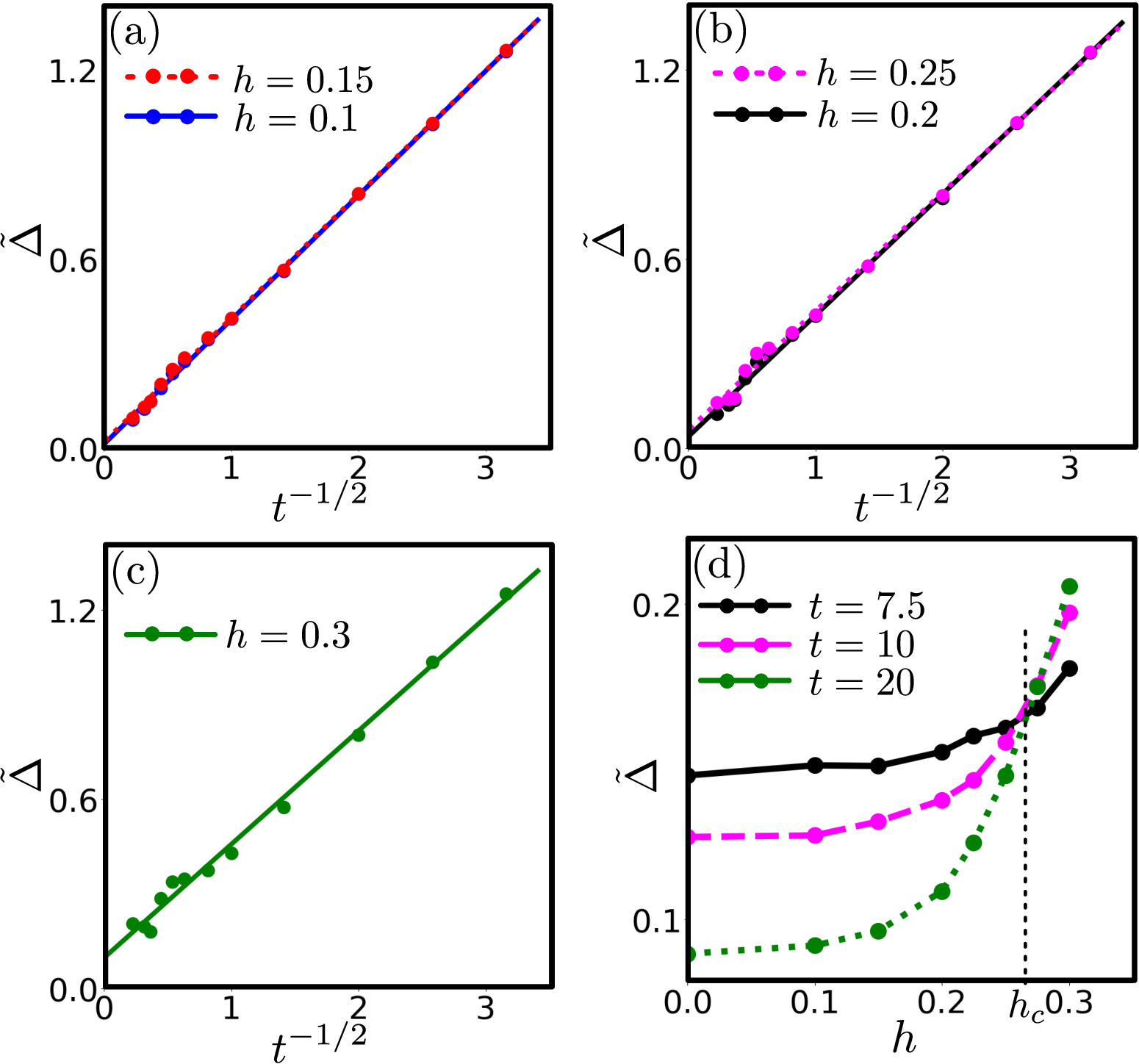}
\caption{(a), (b), and (c) show the residual probability $\tilde{\Delta}$ as a function of $t^{-1/2}$ for different values of $h$ ranging from $h = 0.1$ to $h = 0.3$, distinguished by different colors. Dots represent numerical results, while solid lines indicate fitted curves. To clearly distinguish the different values of $h$, each panel presents a separate subset. 
(d) displays $\tilde{\Delta}$ as a function of $h$ at various times. The curves intersect at approximately $h_c \approx 0.26$, suggesting a crossover behavior. All probability distributions are obtained via $4 \times 10^5$ stochastic samples. The coupling constant $J$ is fixed at $0.2$ throughout.}
\label{fig:delta}
\end{figure}

\subsection{Incomplete SSB in steady-state distribution}

We have demonstrated that the long-time behavior of the probability distribution \( P(\bar{s}) \) is uniquely characterized by the residual probability \( \Delta \). We now numerically investigate \( \Delta \) for \( h \neq 0 \). To do so, we compute the finite-time residual probability, defined as  
$\tilde{\Delta}(t) = {\text{Pr}(-r < \bar{s} < r)}/{\left(2r\right)}$,  
where \( r \) is chosen small enough such that \( \tilde{\Delta}(t) \) accurately approximates the probability density at \( \bar{s} = 0 \). In practice, we find that \( r = 0.05 \) is sufficiently small, as the probability density within the interval \( \bar{s} \in [-0.05, 0.05] \) remains nearly flat for \( t \geq 0.1 \) (see Fig.~\ref{fig:strict}).

For \( h = 0 \), the exact solution yields \( \tilde{\Delta}(t) = 1/\sqrt{2\pi t} \). Thus, we expect a similar \( t^{-1/2} \) decay behavior for small nonzero \( h \). Figures~\ref{fig:delta}(a), (b), and (c) show \( \tilde{\Delta}(t) \) versus \( t^{-1/2} \) for various values of \( h \), with different colors representing different fields. At small \( h \) (e.g., blue dots in panel (a) for \( h = 0.1 \)), we observe an excellent agreement with the \( t^{-1/2} \) decay. We fit the data using the form $\tilde{\Delta}(t) = c t^{-1/2} + \Delta$,
where the fit parameter \( \Delta \) provides an estimate of the long-time residue. For \( h = 0.1 \), the fitted residue is nearly zero (\( \Delta = 0.01 \)). A similarly good fit is observed for \( h = 0.15 \) (red dots), again with negligible residual probability.

As \( h \) increases further (see panels (b) and (c) for \( h = 0.2, 0.25, 0.3 \)), \( \tilde{\Delta}(t) \) exhibits oscillations superimposed on the \( t^{-1/2} \) decay. Nevertheless, the fit form remains useful, as these oscillations tend to average out, allowing us to still extract meaningful values of \( \Delta \). For instance, at \( h = 0.3 \), the asymptotic residue is significant (\( \Delta = 0.1 \)). As \( h \) increases, the residue probability \( \Delta \) continues to rise, while the decay slope \( c \) decreases.

Figure~\ref{fig:delta}(d) plots \( \tilde{\Delta} \) versus \( h \) for several values of \( t \). The intersection of these curves at approximately \( h_c \approx 0.26 \) suggests the presence of a critical field value. Based on this observation, we hypothesize a threshold field \( h_c \) below which \( \Delta = 0 \), and above which the residual probability becomes finite. In this regime, symmetry breaking is incomplete, and a non-negligible probability weight remains near \( \bar{s} = 0 \) even in the long-time limit. While our numerical results strongly support that \( \Delta > 0 \) for sufficiently large \( h \), the existence and precise determination of a sharp transition at finite \( h_c \) remains tentative and should be addressed in future studies using more extensive simulations.

Finally, we emphasize that our RNH Hamiltonian~\eqref{eq:m:mod} with \( \hat{H}_0 = 0 \) describes symmetry-breaking dynamics in the zero-field and zero-temperature limit. When \( \hat{H}_0 \neq 0 \), the RNH Hamiltonian~\eqref{eq:m:mod} predicts a steady state---referred to as an incomplete SSB state---that generally differs from the ground state of \( \hat{H}_0 \). This distinction is evident in the probability distribution \( P(\bar{s}) \), which retains a finite residue probability at \( \bar{s} = 0 \) in contrast to the double-peaked structure expected from the ground state. Consequently, it is not surprising that the observed transition point \( h_c \) differs from the critical field associated with the quantum phase transition of \( \hat{H}_0 \). Developing a more accurate model for real-time dynamics at finite temperature or in the presence of a nonzero field---potentially by modifying the choice of \( \hat{V} \)---remains an important direction for future investigation.

\section{Discussions}
\label{sec:con}

In this paper, we introduce the RNH-Hamiltonian approach 
to model the stochastic nonlinear dynamics of quantum states protected by 
statistical symmetry. Using this framework, we investigate the evolution from 
an initially symmetry-preserved state (paramagnetic) to a SSB state (ferromagnetic). 
Both randomness and non-Hermiticity in the Hamiltonian are essential in capturing this process.

First, multiple SSB branches exist, connected by symmetry transformations. 
As a result, the transition from a unique symmetry-preserving state to the 
SSB states must be a stochastic process. This stochastic process must 
adhere to statistical symmetry, ensuring that different branches of the SSB states 
emerge with equal probability. Traditionally, SSB requires the Hamiltonian to 
have explicit symmetry (in this paper, $\text{Z}_2$). However, we find that, 
to accurately model the dynamics leading to SSB, this explicit symmetry must be 
replaced by statistical symmetry.

Second, the non-Hermitian ($\gamma$) term plays a crucial role in enabling the 
appearance of SSB states during evolution, as it introduces nonlinearity into the 
dynamical equations for physical states. Intuitively, the $\gamma$-term results in 
a non-unitary evolution operator---$e^{\sqrt{\gamma}\hat{\sigma}_z dW_t }$---which 
amplifies the components of the state vector along different $s$-axes in the Hilbert space. 
However, the amplification factor varies along different $s$-axes, and, depending on 
the trajectory of a specific Wiener process, either the $s=1/2$ or $s=-1/2$ axis 
is amplified the most. After normalization, the amplitude of the quantum state 
on $\ket{s=1/2}$ or $\ket{s=-1/2}$ increases, while the amplitudes on all other basis
vectors are suppressed. Over time, the system evolves toward one of the SSB states 
$\ket{s=1/2}$ or $\ket{s=-1/2}$. Superpositions cannot survive this amplification process. 
The probability migration from $\bar{s}=0$ to $\bar{s}=\pm 1/2$ fundamentally relies on 
the magnification and normalization of the state vector's length.

It is important to note that our specific choice of the $\gamma$-term, $i\sqrt{\gamma}\hat{\sigma}_z dW_t$, 
confines the applicability of our model to cases where the final steady states 
are SSB states at zero field and zero temperature. In reality, SSB states at finite 
fields or nonzero temperatures should be $\ket{\pm \bar{s}_T}$, where $0 < \bar{s}_T < 1/2$, 
and each state has a 50\% probability. Our model cannot predict such steady states, 
even with the inclusion of a Hermitian $\hat{H}_0$. To model the dynamics leading to 
these SSB states, a different form of the $\gamma$-term would be required in future work.

Additionally, it is important to emphasize that the dynamics discussed in this paper 
are based on quantum mechanics, distinguishing our approach from those that are 
essentially classical, such as classical nonequilibrium phase transitions~\cite{Henkel08}. 
The RNH-Hamiltonian framework we propose for SSB dynamics is applicable 
to symmetries beyond $\text{Z}_2$. This work provides a general framework for 
describing how a symmetry-preserved quantum state evolves into an SSB state.

\section*{Acknowledgement} 
This work is supported by National Natural Science Foundation of
China (Grants Nos. 11774315, 11835011).

\appendix

\section{Differential equation for normalized physical state}
\label{sec:app:eq}

In this section, we derive the stochastic nonlinear differential equation for the 
normalized state vector using stochastic calculus, a well-established branch of mathematics~\cite{Mikosch98app}.
The model is introduced through its infinitesimal Hamiltonian integral, defined as
$d\hat{H}_t = \hat{H}_0 dt +i \hat{V} dW_t$, where $dW_t$ represents the differential of a Wiener process. 
One might wonder why we do not define the Hamiltonian as $\hat{H}_0 + i\hat{V} \frac{dW_t}{dt}$ but 
instead opt for a Hamiltonian integral, which may seem unfamiliar to the community. Our choice
is motivated by mathematical rigor: the Wiener process is not differentiable, meaning
${dW_t}/{dt}$ does not exist. In statistical mechanics, some authors treat $dW_t/dt$ as a form
of white noise. However, in this paper, we adhere to strict mathematical formalism, thus adopting
the notations of stochastic calculus.

Stochastic calculus differs from ordinary calculus. In stochastic calculus, second-order terms 
cannot simply be discarded; instead, all second-order infinitesimal terms must be carefully considered. 
Besides the first-order term in $dt$, second-order terms like $\left(dW_t\right)^2=dt$ must be retained.
In contrast, other second-order terms such as $dt^2$ or $dW_t dt$ can be ignored.
Similarly, all terms of order $n\geq 3$ can be neglected. Special care is also required when applying
the chain rule of differentiation. When taking the derivative of a composite function, the function
must be expanded into a Taylor series up to the second-order terms, after which each term 
is evaluated to determine whether it should be kept or discarded based on the aforementioned rules.
With this approach, we derive the differential equations for both the prenormalized and normalized state vectors. 

By definition, the infinitesimal evolution operator is $\hat{U}_{dt}= e^{-i d\hat{H}_t}$, which is a nonunitary.
Thus, the prenormalized quantum state after an infinitesimal evolution becomes
$\ket{\phi_{t+dt}} = \hat{U}_{dt} \ket{\phi_{t}} $. 
The differential equation for the prenormalized state is
expressed as
\be\label{eq:s:diff:dphi}
\begin{split}
\ket{d\phi_t} & = e^{-i d\hat{H}_t}\ket{\phi_t} - \ket{\phi_t}  \\ & =
-i d\hat{H}_t \ket{\phi_t} + \frac{1}{2}  \left( -i d\hat{H}_t\right)^2 \ket{\phi_t}
\\ & = -i \hat{H}_0 dt \ket{\phi_t}  +\hat{V} dW_t\ket{\phi_t}
+ \frac{1}{2} \hat{V}^2 dt \ket{\phi_t},
\end{split}
\ee
where we have used the standard rules of stochastic calculus: terms of order \( dt^2 \) and \( dW_t\,dt \) are negligible, while terms of order \( (dW_t)^2 \) must be retained, with \( (dW_t)^2 \) replaced by \( dt \).
The normalized state vector is defined as $\ket{\psi_t}= \ket{\phi_t} /\sqrt{\braket{\phi_t | \phi_t}}$, 
which points in the same direction as $\ket{\phi_t}$ in the Hilbert space, while its length
is normalized to unity. It is straightforward to derive the differential equation for the normalization 
factor $ \braket{\phi_t|\phi_t} $, which reads
\be
\begin{split}
d \braket{\phi_t|\phi_t} & = \braket{\phi_t| d\phi_t}+ \braket{d\phi_t|\phi_t}
+ \braket{d\phi_t|d \phi_t} \\ & = 2 
dW_t\bra{\phi_t} \hat{V} \ket{\phi_t} + 2  dt 
\bra{\phi_t} \hat{V}^2 \ket{\phi_t}.
\end{split}
\ee
Finally, the normalized physical state satisfies
\be
\begin{split}
\ket{ d\psi_t} = & \ d\left( \frac{\ket{\phi_t}}{\sqrt{\braket{\phi_t | \phi_t}}}\right) \\
 =  & -i \hat{H}_0 dt \ket{\psi_t} + 
dW_t \left[ \hat{V} - \langle \hat{V} \rangle\right] \ket{\psi_t} 
\\ & + dt \left\{\frac{1}{2} \left[ \hat{V} - \langle \hat{V} \rangle\right]^2
- \left[ \langle\hat{V}^2 \rangle - \langle \hat{V} \rangle^2 \right] \right\}\ket{\psi_t}.
\end{split}
\ee

\section{Exactly solvable model}
\label{sec:app:st}

The model is exactly solvable when $\hat{H}_0=0$, $\hat{V}=\sqrt{\gamma} \hat{\sigma}_z$, 
and the initial state is $\ket{\psi_0} = \ket{+ + \cdots +}$, which represents a state with all spins 
aligned along the positive $x$-direction. It is convenient to express the state in the Dicke basis, defined as
\be
\ket{s} = \frac{1}{\sqrt{C^n_N}} \sum_{\sigma_1^z+ \cdots+\sigma_N^z=
2Ns}\ket{\sigma_1^z,\sigma_2^z,\cdots,\sigma_N^z},
\ee
where $\sigma_j^z=\pm 1$ indicates whether the $j$-th spin points along the positive or 
negative $z$-direction, respectively. Here, $N$ represents the total number of spins, and 
$n = 0, 1, \cdots, N$ denotes the number of spins aligned along the positive $z$-direction.
$C^n_N$ is the binomial coefficient, and $s= \left(2n-N\right)/2N$ is the average magnetization.
It is easy to see that $\ket{s}$ is the equally-weighted superposition of all spin configurations with 
the same total magnetization. Additionally, $\ket{s}$ is an eigenstate of $\hat{\sigma}_z$ with the 
corresponding eigenvalue of $2Ns$.

\begin{figure}[htp]
\centering
\vspace{0.1cm}
\includegraphics[width=.49\textwidth]{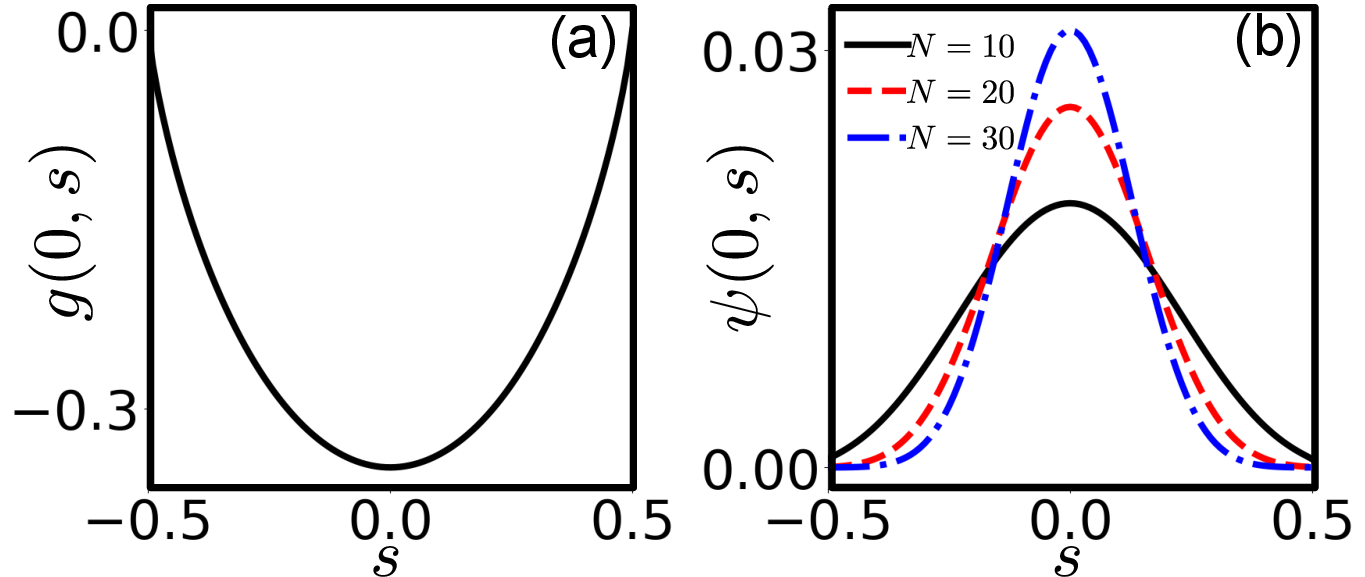}
\caption{(a) Logarithmic wave function at $t=0$. (b) Initial wave function plotted for different values of $N$.}
\label{fig:app:strict}
\end{figure}

In the Dicke basis, the initial wave function is $\psi(0,s) = \braket{s | \psi_0} 
\propto \sqrt{C^n_N}$. Here, we omit any constant factors that are independent of $s$
since the wave function will eventually be normalized at the final time step. Therefore, considering
such factors during intermediate steps is unnecessary. Since $\ket{s}$ is an eigenstate of $\hat{V}$ 
with the eigenvalue $\sqrt{\gamma}2Ns$, we have
\be
\bra{s} e^{- i d\hat{H}_t } = \bra{s} e^{\hat{V} dW_t} = \bra{s}  e^{2Ns\sqrt{\gamma} dW_t} .
\ee
Thus, the wave function at arbitrary time $t$ can be computed as: $\psi(t,s)\propto\phi(t,s)\propto
 e^{2Ns\sqrt{\gamma}W_t} \psi(0,s)$. Next, we analyze the shape of the wave function in $s$-space. 
 Note that $\psi(t,s)$ is real and positive. 

The initial wave function reads
\be\label{eq:m:semi:p0s}
\psi(0,s) \propto \sqrt{\frac{N!}{\left[N(\frac{1}{2}+s)\right]!
\left[N(\frac{1}{2}-s)\right]!}}.
\ee
We are primarily interested in the case where $N$ is sufficiently large. In fact, taking the limit as $N\to\infty$ is essential 
in studying spontaneous symmetry breaking. Here, we assume $N$ is large enough to apply Stirling's approximation.
Using Stirling's formula, we find:
\be
\begin{split}\label{eq:m:str:p00s}
\psi(0,s) \propto \ {\left[\left(\frac{1}{2}+s\right)^{\frac{1}{2}+s}
\left(\frac{1}{2}-s\right)^{\frac{1}{2}-s} \right]^{-\frac{N}{2}}
\left[{2\pi N\left(\frac{1}{4} - s^2\right)}\right]^{-\frac{1}{4}} }.
\end{split}
\ee
From this, it is easy to see that the right-hand side of Eq.~\eqref{eq:m:str:p00s} increases exponentially with $N$.
Thus, without loss of generality, we set $\psi(0,s) \equiv \exp\left\{ - Ng(0,s)\right\}$, where $g$ is called the logarithmic 
wave function. It is straightforward to obtain
\be\label{eq:m:semi:g0}
g(0,s) = \frac{1}{2} \left[
\left(\frac{1}{2}+s\right)\ln\left(\frac{1}{2}+s\right) + 
\left(\frac{1}{2}-s\right)\ln\left(\frac{1}{2}-s\right) \right],
\ee
where we have neglected the $s$-independent constant. In Fig.~\ref{fig:app:strict}(a), we plot $g(0,s)$
as a function of $s$. It is clear that $g$ has a unique minimum. Consequently, we expect $\psi(0,s) = \exp(-Ng(0,s))$ to 
exhibit a single-peak structure, with a peak width on the order of $\sim 1/\sqrt{N}$. As $N$ increases, the peak becomes 
sharper. This is confirmed in Fig.~\ref{fig:app:strict}(b), where plots of normalized $\psi (0,s)$ for different values of $N$ are compared.
In the limit $N\to\infty$, $\psi (0,s)$ approaches a $\delta$-function.

For $t>0$, the wave function in $s$-space can be written as
\be
\begin{split}\label{eq:s:ex:chi}
\psi(t,s) \propto  \ e^{2\sqrt{\gamma} N W_t s} & \left[\left(\frac{1}{2}+s\right)^{\frac{1}{2}+s}
\left(\frac{1}{2}-s\right)^{\frac{1}{2}-s} \right]^{-\frac{N}{2}} \\ & \times
\left[{2\pi N\left(\frac{1}{4} - s^2\right)}\right]^{-\frac{1}{4}} .
\end{split}
\ee
Once again, the right-hand side of Eq.~\eqref{eq:s:ex:chi} increases exponentially with $N$.
We therefore set $\psi(t,s) \equiv \exp\left\{ - Ng(t,s)\right\}$. For sufficiently large $N$, we obtain
\be\label{eq:s:ex:ome}
\begin{split}
g(t,s) = & \ - \frac{1}{N} \ln \psi(t,s) \\ = & \
\frac{1}{2}\left[\left({\frac{1}{2}+s}\right) \ln \left(\frac{1}{2}+s\right)
+\left({\frac{1}{2}-s}\right) \ln \left(\frac{1}{2}-s\right) \right] \\ & \ \ - 2\sqrt{\gamma} s W_t.
\end{split}
\ee
For $t>0$, according to Eq.~\eqref{eq:s:ex:ome}, the curve for $g(t,s)$ can be viewed
as the sum of $g(0,s)$ and a straight line with slope $-4\sqrt{\gamma}W_t$. Adding this straight line to 
Fig.~\ref{fig:app:strict}(a) simply shifts the minimum point to the left if $W_t>0$ or to the right if $W_t<0$,
while the overall shape of $g$ remains the same. As a result, the peak structure of $\psi (t,s)$ is preserved.

To determine the peak location of $\psi (t,s)$, denoted by $\bar{s}$, it is equivalent to find the minimum of $g(t,s)$ 
along the $s$-axis. Solving the equation $\frac{\partial g}{\partial s}\left|_{s=\bar{s}} \right. = 0$, we easily obtain
\be\label{eq:s:ex:bars}
\bar{s} = \frac{1}{2} \tanh\left( \tilde{W}_t\right),
\ee
where $\tilde{W}_t=2\sqrt{\gamma} W_t$, and $W_t$ is a random variable (Wiener process) with a 
Gaussian distribution of zero mean and variance $t$. 
From Eq.~\eqref{eq:s:ex:bars} and probability theory, the distribution of the random variable $\bar{s}$ can be derived. 
The probability density is given by
\be
P(\bar{s}) = \frac{1}{1-4\bar{s}^2} \frac{1}{\sqrt{2\pi\gamma t}}
 \exp \left\{ - \frac{1}{32 \gamma t}
\left( \ln \frac{1+2\bar{s}}{1-2\bar{s}}\right)^2 \right\}.
\ee
Similarly, we can derive the probability density of $\bar{m} = \ln\frac{1+2\bar{s}}{1-2\bar{s}}$,
which is $P(\bar{m})= \frac{1}{\sqrt{32\pi \gamma t}} \exp\left({-\frac{\bar{m}^2}{32\gamma t}}\right)$,
where $\bar{m}$ represents the rescaled magnetization.

At the initial time, the probability of finding $\bar{s}=0$ is $1$. In the limit $t\to\infty$, the probabilities of 
$\bar{s}=\pm 1/2$ become $50\%$ each. This behavior can be understood from the equation 
$\bar{s} = \frac{1}{2} \tanh \left( \tilde{W}_t\right)$, where the variance of $\tilde{W}_t$, equal to $4\gamma t$, diverges
as $t\to\infty$. This means the probability of finding $\tilde{W}_t$ within any finite interval around zero (e.g.,
$\left( \tanh^{-1}\left(-1 + \epsilon \right),\tanh^{-1}\left(1 - \epsilon \right)\right)$) approaches zero. Here, $\epsilon$
denotes an arbitrarily small positive number, and $\tanh^{-1}$ is the inverse hyperbolic tangent function. 
As a result, the probability of $\bar{s}$ lying within $\left( -\frac{1}{2} + \frac{\epsilon}{2}, \frac{1}{2} - 
\frac{\epsilon}{2}\right)$ 
also goes to zero for any $\epsilon>0$. Thus, in the limit, $\bar{s}$ can only take the values $\pm 1/2$.

\section{Stochastic semiclassical approach}
\label{sec:app:ssa}

In this section, we consider the Hamiltonian $\hat{H}_0 = -\frac{J}{N} \hat{\sigma}_z^2 + 
h \hat{\sigma}_x$ and $\hat{V}=\sqrt{\gamma} \hat{\sigma}_z$, and proceed
to solve Eq.~\eqref{eq:s:diff:dphi} to obtain the prenormalized wave function using the stochastic 
semiclassical approach. Normalizing $\ket{\phi_t}$ is straightforward and, in fact, unnecessary, 
as we are only interested in the wave function's properties that are independent of its normalization.

The prenormalized state vector satisfies: $\ket{d\phi_t} = -i\hat{H}_0 dt \ket{\phi_t} +\sqrt{\gamma}
\hat{\sigma}_z dW_t \ket{\phi_t} + \frac{1}{2} \gamma \hat{\sigma}_z^2 dt\ket{\phi_t}$.
Using the Dicke basis, the dynamical equation for the prenormalized wave function becomes:
\be\label{eq:s:sto:dphi}
\begin{split}
d_t \phi(t,s) = & \ i dt \ 4s^2 JN \phi(t,s) -i dt \ N \tilde{f}(s) \cosh \left( \frac{1}{N}\partial_s\right)
\phi(t,s)  \\ & \ + \left[ 2s \sqrt{\gamma} N dW_t  + 2s^2 \gamma N^2
dt  \right] \phi(t,s),
\end{split}
\ee
where $\tilde{f}(s)= 2h\sqrt{\frac{1}{4}-s^2}$ and $d_t \phi(t,s)\equiv \phi(t+dt,s) - \phi(t,s)$. 
Note that $d_t \phi$ is not the total differential, but rather the change in $\phi$ as $t$ changes, 
with $s$ held fixed. It reduces to $dt \partial_t \phi$ in the absence of the Wiener process.
To derive Eq.~\eqref{eq:s:sto:dphi}, we have applied the continuum approximation for the variable $s$, 
a standard technique for fully-connected models in the large $N$ limit~\cite{Sciolla11}. 
In the following, we first discuss the exact solution in the specific case where $h=0$, and then
explore the semiclassical method for $h\neq 0$.

\subsection{Exact solution as $h=0$}
\label{sec:app:seh}

When $ h= 0$ but $\gamma, J\neq 0$, the exact solution for $\phi(t,s)$ can be obtained
since $\ket{s}$ is an eigenstate of $\hat{\sigma}_z$, which simplifies the evolution under $d\hat{H}_t$. 
Similar to the approach in Sec.~\ref{sec:app:st}, the prenormalized wave function is given by:
\be\label{eq:m:semi:phi}
\begin{split}
\phi(t,s) = & \ \bra{s} \exp \left\{ i t \frac{J}{N}\hat{\sigma}_z^2 +\sqrt{\gamma}\hat{\sigma}_z
W_t \right\} \ket{\psi_0} \\  
= & \ \exp\left\{i 4Js^2 Nt +  2Ns \sqrt{\gamma} W_t \right\} \psi(0,s).
\end{split}
\ee
We can rewrite the wave function as $\phi(t,s)= \exp\left\{ - N w(t,s)\right\}$, where
$w$ is the logarithmic wave function, a complex-valued function that can be split into
its real and imaginary components as $w(t,s) = g(t,s) - i\theta(t,s)$.
Using Eqs.~\eqref{eq:m:str:p00s},~\eqref{eq:m:semi:g0}, and~\eqref{eq:m:semi:phi}, we find
\be\label{eq:m:semi:gts}
\begin{split}
& g(t,s) = g(0,s)- 2 \sqrt{\gamma} s W_t , \\
& \theta(t,s)=4Jt s^2.
\end{split}
\ee
Compared to the exact solution when $J=0$, a finite $J$ introduces a nonzero $\theta$, but
does not affect $g$. This exact solution for $h=0$ serves as the foundation for the semiclassical approximation
discussed in the following sections.

\subsection{Semiclassical approximation as $h\neq 0$}

\begin{figure}[htp]
\centering
\vspace{0.1cm}
\includegraphics[width=.48\textwidth]{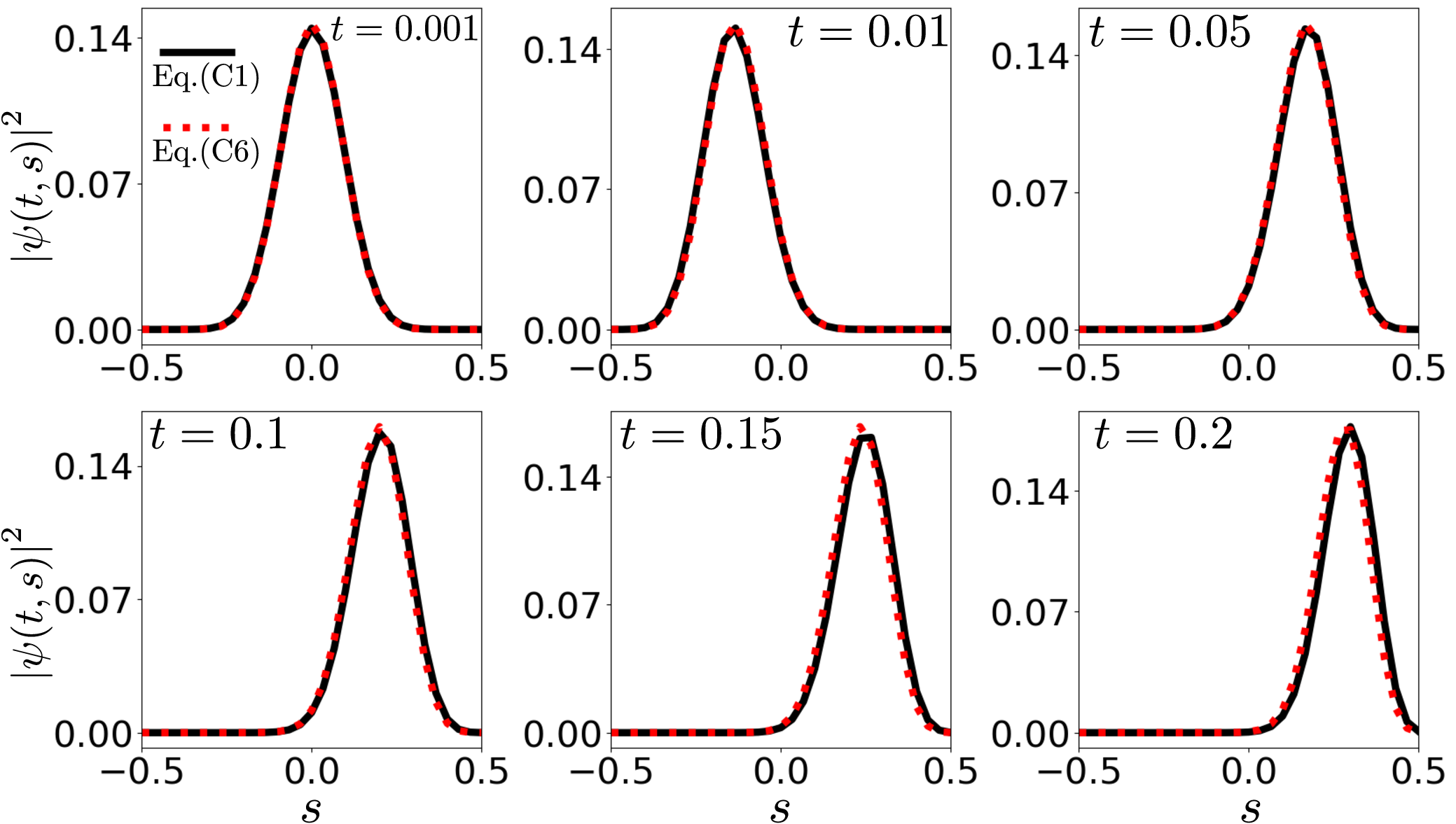}
\caption{Comparison between the exact numerical solution and the approximate one. The black solid line shows the result obtained by directly evolving the original stochastic Schrödinger equation (Eq.~\eqref{eq:s:sto:dphi}) for a system of \( N = 30 \) spins. The red dotted line corresponds to the solution of the approximate equation (Eq.~\eqref{eq:gtsemieqwithJ}), which neglects terms of order \( \mathcal{O}(1/N) \). A clear shift of the peak away from the origin reveals that \( \bar{s} \neq 0 \), indicating spontaneous symmetry breaking. Parameters are set to \( J = 0.1 \), \( h = 0.2 \), and \( \gamma = 1 \). Different panels correspond to different evolution times.}
\label{fig:app:ch16}
\end{figure}

As $h\neq 0$, we develop the stochastic semiclassical approach, which
relies on the fact that the wave function (whether prenormalized or
normalized) exhibits a single-peak structure in $s$-space, with a peak width of approximately $1/\sqrt{N}$. 
As a result, instead of solving for the entire wave function, we focus on identifying the position of the peak. 
The stochastic semiclassical method systematically derives the equation governing the peak's position.
First, we substitute $\phi \equiv e^{-Nw}$ into Eq.~\eqref{eq:s:sto:dphi}. Using the approximation
\be
 \left( \frac{1}{N}\partial_s\right)^n e^{ -N w (t,s)} = e^{ -N w (t,s)}
 \left[ \left(-\partial_s w\right)^n +  \mathcal{O}(\frac{1}{N})\right],
\ee
and neglecting the $ \mathcal{O}(\frac{1}{N})$ terms, we obtain the following equation for 
the logarithmic wave function $w$:
\be\label{eq:m:semi:dtw}
d_t w = i dt \  \tilde{f}(s) \cosh (\partial_s w) - i dt \ 4s^2 J - 2s \sqrt{\gamma} dW_t .
\ee
By decomposing $w$ into its real and imaginary parts, $w(t,s)=g(t,s) - i\theta(t,s) $, we
find that Eq.~\eqref{eq:m:semi:dtw} becomes equivalent to
\be\label{eq:gtsemieqwithJ}
\begin{split}
& d_t g = dt \tilde{f} \sin(\partial_s\theta) \sinh(\partial_s g) -  2s  \sqrt{\gamma} \ {dW_t}, \\
& d_t\theta = - dt \tilde{f} \cos(\partial_s\theta) \cosh(\partial_s g) + dt \ 4s^2 J.
\end{split}
\ee
From Eq.~\eqref{eq:s:sto:dphi} to Eq.~\eqref{eq:gtsemieqwithJ}, we neglect the $ \mathcal{O}(\frac{1}{N})$ terms, 
a valid approximation for large $N$, as demonstrated in previous studies where $\gamma=0$~\cite{Sciolla11}, 
i.e., in the absence of randomness and nonHermiticity. To verify the validity of Eq.~\eqref{eq:gtsemieqwithJ}
for $\gamma\neq 0$, we numerically solve the original equation, Eq.~\eqref{eq:s:sto:dphi}, and compare
its solution with Eq.~\eqref{eq:gtsemieqwithJ}. Figure~\ref{fig:app:ch16} shows a comparison for \( N = 30 \), across a sequence of times from \( \gamma t = 0.001 \) to \( \gamma t = 0.2 \). The two solutions---obtained from Eq.~\eqref{eq:s:sto:dphi} (black solid lines) and Eq.~\eqref{eq:gtsemieqwithJ} (red dotted lines)---are nearly indistinguishable, indicating that \( N = 30 \) is sufficiently large for the approximation in Eq.~\eqref{eq:gtsemieqwithJ} to be valid at all times. This observation is consistent with previous insights into semiclassical approximations~\cite{Sciolla11}.
We emphasize, however, that obtaining accurate numerical solutions to the nonlinear equation~\eqref{eq:gtsemieqwithJ} requires a very small time step. This is due to the rapidly growing numerical error in evaluating the hyperbolic functions \( \sinh(\partial_s g) \) and \( \cosh(\partial_s g) \), especially when \( g(t, s) \) varies sharply with \( s \), which typically occurs near the endpoints \( s = \pm 1/2 \). As a result, the numerical integration of Eq.~\eqref{eq:gtsemieqwithJ} is generally limited to early and intermediate times.

To solve Eq.~\eqref{eq:gtsemieqwithJ}, we note that $\left| \phi\right| = e^{-Ng} $ has a single-peak 
structure, implying that $g$ must have a unique minimum in $s$-space (see Fig.~\ref{fig:app:strict}(a)
for $g(0,s)$). We denote this minimum point as $\bar{s}$, which corresponds to the peak position
of $\left| \phi\right|$ and varies with time. Around $\bar{s}(t)$, we can expand $g(t,s)$ and $\theta(t,s)$
into Taylor's series:
\be
\begin{split}\label{eq:m:semi:gtset}
 g(t,s) = & \ g_0 (t) + \frac{1}{2 !} g_2(t) \left[ s -\bar{s}(t) \right]^2 + \cdots+
\frac{1}{n !} g_n (t) \left[ s -\bar{s}(t) \right]^n \\ & + \cdots, \\
 \theta(t,s) = & \ \theta_0 (t) + \theta_1 (t) \left[ s -\bar{s}(t) \right] + \cdots+
\frac{1}{n !} \theta_n (t) \left[ s -\bar{s}(t) \right]^n \\ & + \cdots.
\end{split}
\ee
Since $\partial_s g = 0 $ at $ s=\bar{s}$, there is no first-order term in the expansion of $g$.
Meanwhile, $\theta_1(t)= \left. \partial_s\theta(t,s) \right|_{s=\bar{s}}$, often referred to as the classical momentum
and denoted as $\bar{p}(t)$ in semiclassical literature. 
Using the expansion in Eq.~\eqref{eq:m:semi:gtset}, we derive the differential equations
for $g$ and $\theta$ as functions of time, while keeping $s$ fixed:
\begin{widetext}
\be\label{eq:m:semi:dgdtas}
\begin{split}
d_t g  = & \ \left\{d{g}_0+\frac{1}{2}g_2 \left(d\bar{s}\right)^2\right\}  +
\left\{   - g_2 d\bar{s} -dg_2 d\bar{s} + \frac{1}{2}g_3\left(d\bar{s}\right)^2 \right\} 
\left[ s -\bar{s} \right] + \cdots \\ & +
\left\{ \frac{1}{n !} dg_n - \frac{1}{n !} g_{n+1} d\bar{s}-\frac{1}{n !}
dg_{n+1} d\bar{s} + \frac{1}{n !} \frac{1}{2} g_{n+2} \left(d\bar{s}\right)^2  
\right\} \left[ s -\bar{s} \right]^n  + \cdots, \\
d_t \theta = & \ \left\{ d\theta_0 -\theta_1 d\bar{s} -d\theta_1 d\bar{s}
+\frac{1}{2}\theta_2 \left(d\bar{s}\right)^2 \right\}  +  \left\{ 
d\theta_1 -\theta_2 d\bar{s} -d\theta_2 d\bar{s}
+\frac{1}{2}\theta_3 \left(d\bar{s}\right)^2 \right\} \left[ s -\bar{s} \right]  + \cdots \\ & +
\left\{ \frac{1}{n !} d\theta_n - \frac{1}{n !} \theta_{n+1} d\bar{s}-\frac{1}{n !}
d\theta_{n+1} d\bar{s} + \frac{1}{n !} \frac{1}{2} \theta_{n+2} \left(d\bar{s}\right)^2  
\right\} \left[ s -\bar{s} \right]^n  + \cdots.
\end{split}
\ee
\end{widetext}
Substituting these into the left-hand side of Eq.~\eqref{eq:gtsemieqwithJ} and similarly expanding
the right-hand side, we arrive at a series of equations for $\bar{s}$, $\bar{p}$, $g_n$ and $\theta_n$.
The first-order expansion yields the critical equation:
\be
\begin{split}\label{eq:semieqforspJ}
 -g_2 d\bar{s}-dg_2 d\bar{s}+\frac{1}{2}g_3\left(d\bar{s}\right)^2 =& \ dt \tilde{f}_0 g_2 \sin(\bar{p}) 
 -2\sqrt{\gamma} dW_t , \\ 
-d \bar{p} + \theta_2 d\bar{s} + d\theta_2 d\bar{s}-\frac{1}{2} \theta_3
\left(d\bar{s}\right)^2  =& \ dt\left\{ \tilde{f}_1\cos(\bar{p})-\tilde{f}_0 \theta_2 \sin(\bar{p})\right\}
\\ & \ -dt 8J\bar{s},
\end{split}
\ee
where the Taylor's coefficients $\tilde{f}_n$ for $\tilde{f}(s)$ around $s=\bar{s}$ are given by:
\be
\begin{split}
\tilde{f}_0 = 2h \sqrt{\frac{1}{4}-\bar{s}^2}, \ \ 
\tilde{f}_1 = \frac{-2h\bar{s}}{\sqrt{\frac{1}{4}-\bar{s}^2}}, \ \
\tilde{f}_2 =- \frac{h}{2}\frac{1}{{\sqrt{\frac{1}{4}-\bar{s}^2}}^3}.
\end{split}
\ee
This system allows us to compute the peak position of the wave function.
In the limit $N\to\infty$, the wave packet shrinks to a $\delta$-function, making the peak position
the most significant feature of $\phi$. Other features become negligible for sufficiently large $N$.

Equation~\eqref{eq:semieqforspJ} contains several unknown Taylor coefficients, namely $g_2$, 
$g_3$, $\theta_2$ and $\theta_3$. When $\gamma=0$, these unknowns cancel each other out,
leading to self-consistent equations for $\bar{s}$ and $\bar{p}$. This explains why the semiclassical 
theory becomes exact as $N\to\infty$ in models with Hermitian Hamiltonians. However, when
nonHermiticity is present (i.e., $\gamma\neq 0$), such cancellations do not occur.
Moreover, including higher-order terms in the expansion does not resolve the issue, 
as no self-consistent set of equations for $\bar{s}$, $\bar{p}$, $g_n$ and $\theta_n$ can be
obtained at any order of truncation.

To overcome this problem, we use the exact solution for $h=0$ (see Sec.~\ref{sec:app:seh}), 
where the transverse field is absent. For $J,\gamma\neq 0$ but $h=0$, we can already obtain 
the exact expression for the wave function $\phi$ as well as the logarithmic functions $g(t,s)$ and $\theta(t,s)$.
Using their expressions in Eq.~\eqref{eq:m:semi:gts}, we can compute the corresponding Taylor series, 
with the coefficients given by
\be
\begin{split}\label{eq:g2g3J0}
& g_2(t) = \frac{2}{1- \tanh^2\left(\tilde{W}_t\right)}, \ \ 
g_3(t) = \frac{8\tanh\left(\tilde{W}_t\right)}{\left[1- \tanh^2\left(\tilde{W}_t\right)\right]^2}, \\ & 
\theta_2= 8Jt, \ \ \theta_3=0,
\end{split}
\ee
where we have used the notation $\tilde{W}_t =2 \sqrt{\gamma} W_t$.
Next, we consider the case where $h$ is small but nonzero. We approximate that 
the expressions for $g_n$ and $\theta_n$ obtained for $h=0$ remain valid for small values of $h$. 
Substituting Eq.~\eqref{eq:g2g3J0} into Eq.~\eqref{eq:semieqforspJ}, and applying techniques from stochastic calculus,
we simplify the equations for $\bar{s}$ and $\bar{p}$ as follows:
\be\label{eq:dsdpeqh0J}
\begin{split}
& d\bar{s} = d\left(\frac{1}{2}\tanh\left(\tilde{W}_t\right)\right)
-2h\sqrt{\frac{1}{4}-\bar{s}^2} \sin\left(\bar{p}\right) dt, \\ &
d\bar{p} = 2h \frac{\bar{s}}{\sqrt{\frac{1}{4}-\bar{s}^2}} \cos(\bar{p}) dt
+ 4Jt d\left( \tanh\left(\tilde{W}_t\right)\right) + 8J\bar{s} dt.
\end{split}
\ee
Note that Eq.~\eqref{eq:dsdpeqh0J} describes a non-Markovian process.
This is because $d\left( \tanh\left(\tilde{W}_t\right)\right)$ differs from $d {W}_t$; while the latter
is, by definition, a random variable independent of previous values due to the independent increment 
property of the Wiener process, $d\left( \tanh\left(\tilde{W}_t\right)\right)$ depends on $\tanh\left(\tilde{W}_t\right)$, 
which is influenced by the history of increments ($dW_t$) before time $t$.
Consequently, no Fokker-Planck equation can be derived for Eq.~\eqref{eq:dsdpeqh0J} due to
its non-Markovian nature.

\begin{figure}[htp]
\centering
\vspace{0.1cm}
\includegraphics[width=.49\textwidth]{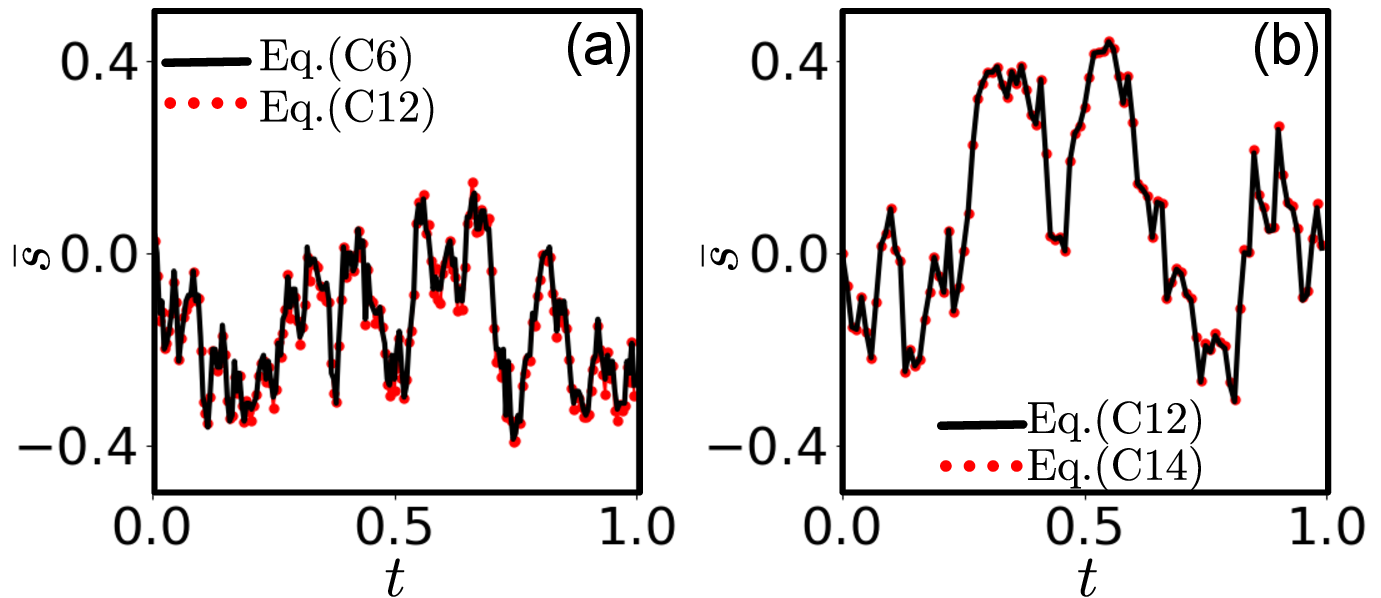}
\caption{Comparison between the trajectories of \( \bar{s}(t) \) obtained from different approximation schemes. The parameters are set to \( J = 0.1 \), \( h = 0.2 \), and \( \gamma = 1 \). 
(a) The black solid line is obtained by solving Eq.~\eqref{eq:gtsemieqwithJ} and identifying the minimum of the wave function, while the red dotted line corresponds to the trajectory computed from Eq.~\eqref{eq:dsdpeqh0J}.
(b) The black solid line shows the solution to Eq.~\eqref{eq:dsdpeqh0J}, and the red dotted line represents the first-order perturbative result from Eq.~\eqref{eq:m:semi:fsp}.
Note that the stochastic dynamics are simulated using a pseudo-random number generator, so the specific trajectories may vary across different runs.}
\label{fig:app:che}
\end{figure}

To validate the use of $g_n$ and $\theta_n$ for $h=0$, we numerically solve 
Eq.~\eqref{eq:gtsemieqwithJ} (which has already been shown to be effective)
and locate the minimum point of $g$ in $s$-space. The corresponding trajectory of the minimum point 
is shown as the black solid line in Fig.~\ref{fig:app:che}(a). For comparison, we also numerically evolve Eq.~\eqref{eq:dsdpeqh0J} to obtain $\bar{s}(t)$, which is plotted as red dots in Fig.~\ref{fig:app:che}(a). 
The excellent agreement between the two indicates that our approximation works well for small values of $h$, such as $h=0.2$.

In addition to the numerical solution, we can also solve Eq.~\eqref{eq:dsdpeqh0J}
analytically using perturbation theory, expressing the solution as a power series in $h$. 
The zeroth-order solution, obtained by setting $h=0$, is given by:
\be\label{eq:solspJ}
\begin{split}
\bar{s}^{(0)}(t) = & \ \frac{1}{2} \tanh \left(\tilde{W}_t\right), \\
\bar{p}^{(0)}(t) =  & \ 4Jt \tanh \left(\tilde{W}_t\right) .
\end{split}
\ee
Substituting Eq.~\eqref{eq:solspJ} into the right-hand side of Eq.~\eqref{eq:dsdpeqh0J} and integrating over $t$, 
we can obtain the first-order solution for $\bar{s}$ and $\bar{p}$, which reads:
\be\label{eq:m:semi:fsp}
\begin{split}
 \bar{s} (t) \approx & \ \frac{1}{2}\tanh\left(\tilde{W}_t\right)-h \int^t_0
d\tau \sqrt{1-\tanh^2\left(\tilde{W}_\tau \right)} \\ & \ \ \ \ \ \ \ \ \ \ \ \ \ \ \ 
\ \ \ \ \ \ \ \ \ \ \ \ \ \ \times \sin
\left( 4J\tau \tanh\left(\tilde{W}_\tau\right) \right), \\
 \bar{p}(t) \approx & \ 4J  t \tanh\left(\tilde{W}_t\right) + 2h \int^t_0 d\tau \frac{\tanh\left(\tilde{W}_\tau\right) }
{\sqrt{ 1- \tanh^2 \left(\tilde{W}_\tau\right)}}  \\ & \ \ \ \ \ \ \ \ \ \ \ \ \ \ \ 
\ \ \ \ \ \ \ \ \ \ \ \ \ \ \times \cos \left( 4J \tau\tanh \left(\tilde{W}_\tau \right)\right).
\end{split}
\ee
This process can be continued to obtain the solution to any order in $h$.
In practice, we find the first-order solution is sufficiently accurate for the
parameter range we are interested in. In Fig.~\ref{fig:app:che}(b), we compare
the solution for $\bar{s}(t)$ obtained from directly evolving Eq.~\eqref{eq:dsdpeqh0J}
with the first-order solution from Eq.~\eqref{eq:m:semi:fsp}. The excellent agreement demonstrates
the validity of the first-order solution.


\section{Real-time dynamics of $\bar{s}$}
\label{sec:app:rts}

In the previous section, we used the stochastic semiclassical approach to solve
the dynamical equation of the wave function and determine its peak position, $\bar{s}$.
We demonstrated that the solution in Eq.~\eqref{eq:m:semi:fsp} provides a good approximation.
Next, we will examine the real-time dynamics of $\bar{s}$ using Eq.~\eqref{eq:m:semi:fsp}.

\begin{figure*}[htp]
\centering
\vspace{0.1cm}
\includegraphics[width=.75\textwidth]{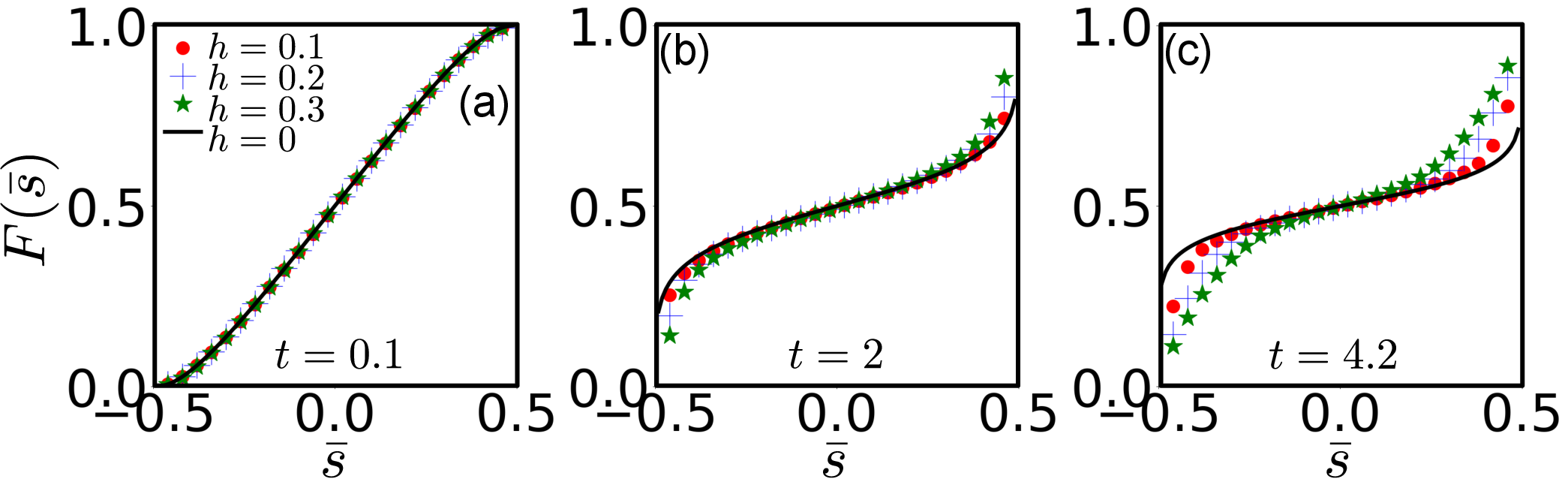}
\caption{Cumulant distribution functions $F(\bar{s})$ at (a) $t=0.1$, (b) $t=2$, and 
(c) $t=4.2$, obtained by using Eq.~\eqref{eq:m:semi:fsp}. The parameter $J$ is set to 0.1. Results for different values of $h$ are represented using lines and dots of varying types and colors. Each distribution is 
based on $10^4$ samples to ensure statistical accuracy.}
\label{fig:app:sdh}
\end{figure*}

Recall that $\tilde{W}_\tau = 2\sqrt{\gamma} W_\tau$ ($0\leq \tau \leq t$)
is a random variable, and the properties of $W_\tau$ are well-studied in stochastic calculus. 
In numerical simulations, $W_\tau$ can be generated using $W_\tau = \Delta W_{t_0}+ 
\Delta W_{t_1}+\cdots + \Delta W_{t_{n-1}}$, where $\tau = n \Delta t$ and $\Delta t$ is 
the time step, chosen to be sufficiently small. The terms $\Delta W_{t_j}$, with $j=0,1,\cdots, n-1$, 
are independent Gaussian random variables, each with a mean of zero and a variance of $\Delta t$.
For a given time $t$, the stochastic process $W_\tau$ over $0\leq \tau \leq t$
is not repeatable in each simulation. Since $\bar{s}$ is expressed 
in terms of $\tilde{W}_\tau$, it is also not repeatable. Therefore, we need to repeat the simulation
multiple times to capture the statistical properties, such as the cumulant
distribution of the random variable $\bar{s}$. In practice, we find that $10^4$ samples
are sufficient to achieve statistical convergence.

Figure~\ref{fig:app:sdh} shows the cumulant distribution function
of $\bar{s}$ at various times and for different values of $h$, with $J=0.1$ fixed.
The exact solution for $h=0$ is also plotted for comparison.
At early times (e.g., $t=0.1$), the distribution for finite $h$ ($h=0.1$, $0.2$, or $0.3$)
shows no significant deviation from the distribution at $h=0$. However, as time increases,
the deviation gradually becomes noticeable, with the extent of deviation
depending on the value of $h$. By $t=2$, $F(\bar{s})$ for $h=0.3$ shows
a clear deviation from the exact solution (see the green stars in Fig.~\ref{fig:app:sdh}(b)),
but $F(\bar{s})$ for $h=0.1$ remains close to the exact solution, indicating that
the deviation appears earlier as $h$ increases.
At a later time ($t=4.2$), the deviations for $h=0.1$, $0.2$, and $0.3$ become more pronounced,
with the magnitude of deviation increasing with $h$. At the same time,
we observe that the slope of $F(\bar{s})$ near $\bar{s}=0$ decreases with time,
and the deviation from the exact solution causes the slope to decrease more gradually.
Since the slope of $F(\bar{s})$ represents the probability density, we conclude
that  the probability migration from $\bar{s}=0$ to $\bar{s}=\pm 1/2$ slows down as $h$ increases. 
In other words, the transverse field slows the probability migration.

\section{Steady-state distribution of $\bar{s}$}
\label{sec:app:ssds}

In the previous section, we discussed the evolution of the distribution of $\bar{s}$ 
using the first-order perturbative solution in Eq.~\eqref{eq:m:semi:fsp}. However, 
Eq.~\eqref{eq:m:semi:fsp} is valid only for short to intermediate times. If $t$ becomes 
too large, the equation may result in nonphysical values of $\left|\bar{s}\right| > 1/2$. 
As $t \to \infty$, we expect the distribution of $\bar{s}$ to relax into a steady state. 
For $h=0$, this steady state corresponds to $\bar{s} = \pm 1/2$, with each value 
having a probability of 50\%. For $h > 0$, we must revisit the nonlinear stochastic 
equation~\eqref{eq:dsdpeqh0J} to explore the steady-state distribution.

In Eq.~\eqref{eq:dsdpeqh0J}, as $t \to \infty$, the variance of $\tilde{W}_t$, which is $4\gamma t$, 
diverges. From probability theory, it is clear that $\tanh(\tilde{W}_t) \to \pm 1$ in this 
limit. More precisely, the random variable $\tanh(\tilde{W}_t)$ approaches $\pm 1$ 
with 100\% probability, and the probability of it taking a value within the interval 
$(-1 + \epsilon, 1 - \epsilon)$, for any arbitrarily small $\epsilon > 0$, becomes zero 
as $t \to \infty$. Since $\tanh(\tilde{W}_t) \to \pm 1$, we also have $d(\tanh(\tilde{W}_t)) \to 0$ 
with 100\% probability. Thus, the dynamical equations~\eqref{eq:dsdpeqh0J} reduce to
\begin{equation}
\begin{split}\label{eq:m:ste:dsdp}
& d\bar{s} = 
-2h\sqrt{\frac{1}{4}-\bar{s}^2} \sin\left(\bar{p}\right) dt, \\ &
d\bar{p} = 2h \frac{\bar{s}}{\sqrt{\frac{1}{4}-\bar{s}^2}} \cos(\bar{p}) dt + 8J\bar{s} dt.
\end{split}
\end{equation}
In the limit $t \to \infty$, only the terms involving $J$ and $h$ remain, while the terms 
related to $\gamma$ vanish. This indicates that the non-Hermitian random part of 
the Hamiltonian loses its influence on $\bar{s}$, the peak position of the wave function.

Equation~\eqref{eq:m:ste:dsdp} is identical to the one found in the semiclassical 
theory of the transverse-field Ising model~\cite{Sciolla11}. This can also be 
seen within our stochastic semiclassical approach by setting $\gamma = 0$ in 
Eq.~\eqref{eq:semieqforspJ}. With $\gamma=0$, the random terms involving 
$dW_t$ are removed from Eq.~\eqref{eq:semieqforspJ}, and the second-order differentials 
disappear, as all the first-order differentials are now proportional to $dt$. As a result,
the terms $g_n$ and $\theta_n$ in Eq.~\eqref{eq:semieqforspJ} cancel each other,
leading to the emergence of Eq~\eqref{eq:m:ste:dsdp}. If we consider the quantum
dynamics governed by the Hermitian Hamiltonian 
$\hat{H}_0=-\frac{J}{N}\hat{\sigma}_z^2 + h\hat{\sigma}_x $, with the initial
state having all spins aligned along the positive $x$-direction, the quantities
$\bar{s}$ and $\bar{p}$ obtained from the wave packet strictly satisfies
Eq.~\eqref{eq:m:ste:dsdp}.

\begin{figure}[htp]
\centering
\vspace{0.1cm}
\includegraphics[width=.3\textwidth]{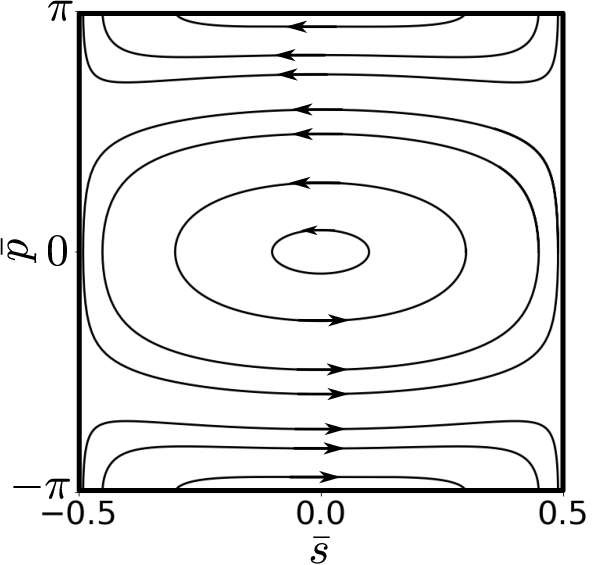}
\caption{Equal-energy contours in the $\bar{s}-\bar{p}$ plane, governed by
the classical dynamics described by the Hamiltonian $H (\bar{s},\bar{p})= 
2h \sqrt{\frac{1}{4}-\bar{s}^2} \cos \bar{p} - 4J\bar{s}^2$. Arrows indicate the direction
of motion along the contours. Parameters are set to $h=0.2$ and $J=0.1$. }
\label{fig:app:cla}
\end{figure}
Moreover, Eq.~\eqref{eq:m:ste:dsdp} describes the motion of a classical point in
the $\bar{s}-\bar{p}$ plane, governed by classical Hamiltonian equations:
$d\bar{p} /dt = \partial H/\partial \bar{s} $ and $d\bar{s} /dt =- \partial H/\partial \bar{p} $,
where the Hamiltonian is given by $H (\bar{s},\bar{p})= 2h \sqrt{\frac{1}{4}-\bar{s}^2}
 \cos \bar{p} - 4J\bar{s}^2$. The Hamiltonian $H$ is conserved, remaining invariant 
throughout the system's evolution. In Fig.~\ref{fig:app:cla}, we show an example of 
constant $H$ contours (equal-$H$ lines) in the $\bar{s}-\bar{p}$ plane,
with parameters set to $h=0.2$ and $J=0.1$. The arrows indicate the direction of 
motion for the point $\left(\bar{s},\bar{p}\right)$ along these contours.
In the $\bar{s}-\bar{p}$ plane, $d\bar{s}/dt$ and $d\bar{p}/dt$ can be interpreted as
the velocity of a fluid. This allows us to derive a corresponding Fokker-Planck equation,
as the dynamics described by Eq.~\eqref{eq:m:ste:dsdp} now represent a Markovian
process, with the nonMarkovian terms such as $d\tanh\left(\tilde{W}_t\right)$
having been neglected. The Fokker-Planck equation is given by:
\be
\begin{split}
\partial_t P(\bar{s},\bar{p},t) = & \ \partial_{\bar{s}}
\left[2h\sqrt{\frac{1}{4}-\bar{s}^2} \sin\left(\bar{p}\right) \cdot P(\bar{s},\bar{p},t) \right] 
\\ & \ - \partial_{\bar{p}} \left[ \left(2h \frac{\bar{s}}{\sqrt{\frac{1}{4}-\bar{s}^2}} \cos(\bar{p}) + 8J\bar{s}
\right) \cdot P(\bar{s},\bar{p},t)  \right].
\end{split}
\ee


\end{document}